%% file: Jarvis-HEP-new.tex
\documentclass[preprint,5p,twocolumn,times,sort&compress]{elsarticle}
\journal{Computer Physics Communications}
\include{def}

\begin{document}
\begin{frontmatter}

\title{\includegraphics[width=64pt]{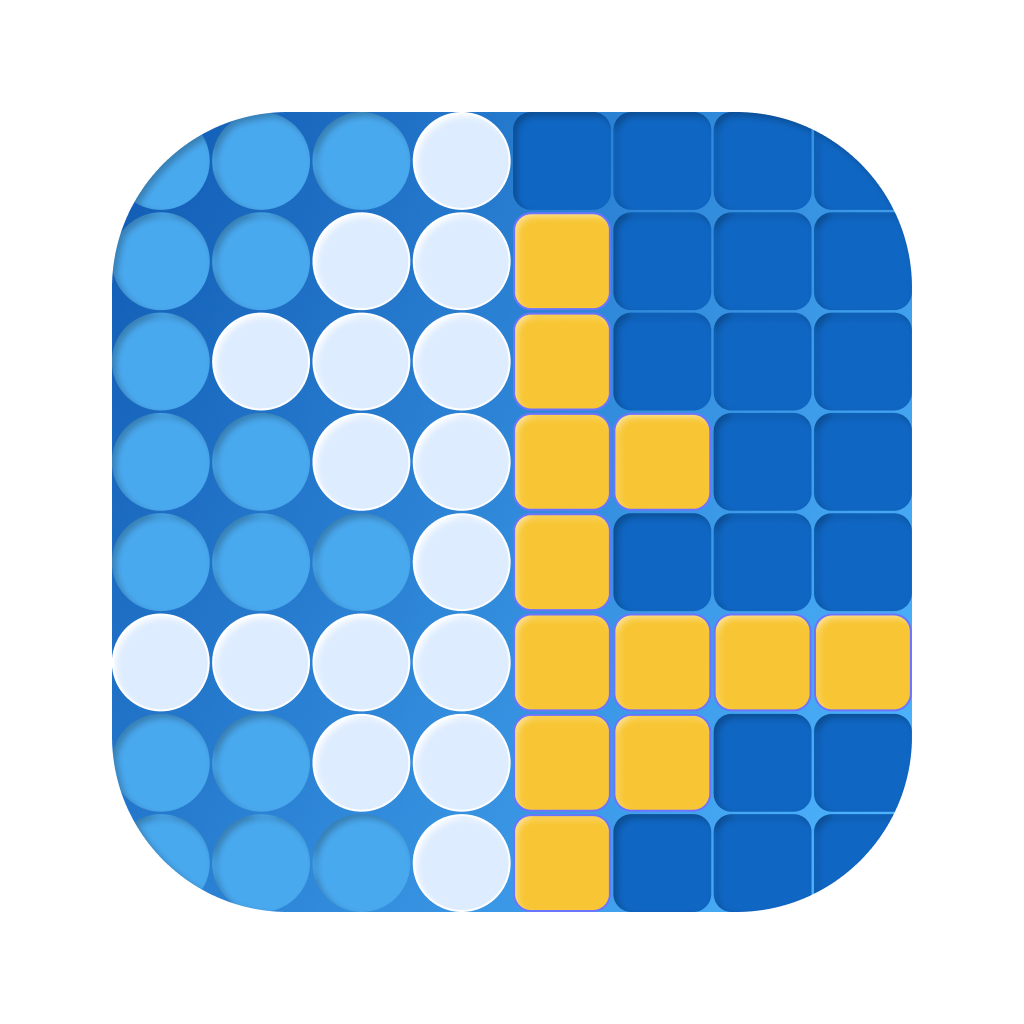} \\\code{Jarvis-HEP}: A lightweight Python framework for workflow composition and parameter scans in high-energy physics
}

\author[a,b]{Erdong Guo}
\author[c]{Paul Jackson}
\author[d,e]{Jin Min Yang}
\author[c]{Pengxuan Zhu}

\cortext[author]{ 
\textit{E-mail address:}  \\
\href{mailto:eguo1@ucsc.edu}{eguo1@ucsc.edu} (E. Guo), \\
\href{mailto:p.jackson@adelaide.edu.au}{p.jackson@adelaide.edu.au} (P. Jackson), \\
\href{mailto:jmyang@itp.ac.cn}{jmyang@itp.ac.cn}(J. M. Yang), \\
\href{mailto:pengxuan.zhu@adelaide.edu.au}{pengxuan.zhu@adelaide.edu.au} (P. Zhu).}

\address[a]{University of California, Santa Cruz, 1156 High St, Santa Cruz, CA 95064, United States}
\address[b]{University College London, Gower St, London WC1E 6BT, United Kingdom}
\address[c]{ARC Centre of Excellence for Dark Matter Particle Physics, \\
Department of Physics, Adelaide University, Adelaide, SA 5005, Australia}
\address[d]{Institute of Theoretical Physics, Chinese Academy of Sciences, Beijing 100190, China}
\address[e]{Centre for Theoretical Physics, Henan Normal University, Xinxiang 453007, China}

\begin{abstract}
High-energy physics phenomenology often requires linking multiple computational tools to evaluate observables, likelihoods, and experimental constraints across nontrivial parameter spaces. 
In this work, we introduce \code{Jarvis-HEP}, a lightweight Python framework for workflow composition and parameter scans in high-energy physics.
The framework provides YAML-based workflow specification, dependency-aware execution, modular calculator integration, and asynchronous task scheduling for multi-step computational studies. 
It supports both external software packages and internally implemented components within a unified workflow, and the current implementation includes several built-in sampling backends for exploratory scans. 
This paper describes the design and user interface of \code{Jarvis-HEP} and illustrates its use with representative synthetic and phenomenological examples.
\\

\noindent \textbf{PROGRAM SUMMARY}  
\noindent \\
\begin{small}
{\em Program Title:} \code{Jarvis-HEP} \\
{\em CPC Library link to program files:} (to be added by Technical Editor) \\
{\em Developer's repository link:} \url{https://github.com/Pengxuan-Zhu-Phys/Jarvis-HEP} \\
{\em Licensing provisions:} MIT  \\
{\em Programming language:} \code{Python 3}                                   \\
{\em Nature of problem:} \\
High-energy physics phenomenology often requires combining multiple computational tools in order to evaluate observables, likelihoods, and experimental constraints across nontrivial parameter spaces. In practice, these studies involve multi-step workflows, external software integration, and consistent handling of dependencies during parameter scans. Managing such workflows manually can be cumbersome and error-prone, especially when calculations must be repeated over many parameter points.
\\
{\em Solution method:} \\
\code{Jarvis-HEP} is a lightweight Python framework for workflow composition and parameter scans in high-energy physics. It provides YAML-based workflow specification, dependency-aware execution, modular calculator integration, and asynchronous task scheduling for multi-step computational studies. The framework supports both external software packages and internally implemented components within a unified workflow, and the current implementation includes several built-in sampling backends together with monitoring and logging utilities.
\end{small}
\end{abstract}
\end{frontmatter}
\clearpage
\tableofcontents

\section{Introduction}
\label{sec:int}
The search for new physics beyond the Standard Model (BSM) of particle physics has been fueled by ever-larger collider experiments, dark matter (DM) experiments, astronomy and cosmology. 
Large observational datasets are used to test increasingly sophisticated theoretical models.
Performing a phenomenological study requires comparing theoretical predictions of BSM models with experimental data using a variety of computational tools. 
This software ecosystem has been developed through broad community effort. 
At the same time, BSM phenomenology has become computationally intensive and methodologically complex: large parameter spaces must be explored, multiple calculations must be coordinated, and predictions must be compared with experimental data in a consistent manner.

Since the early 2000s, supersymmetry (SUSY), especially the Minimal Supersymmetric Standard Model (MSSM), has played a leading role in standardizing data interface through the SUSY Les Houches Accord (SLHA) format~\cite{Skands:2003cj, Allanach:2008qq}. 
The SLHA effectively links the codes involved in SUSY research. 
The success of this standardization provides a paradigm for the numerical studies of other new physical theories. 
Alongside these interface-standardization efforts, the software ecosystem for phenomenological studies has expanded substantially.
At present, the various packages for new-physics studies fall into several broad categories, depending on their primary functions:
\begin{itemize}
	\item Spectrum generators, which calculate the particle masses and couplings within a candidate BSM model, including running effects described by the renormalisation group equations~\cite{Staub:2008uz, Staub:2013tta, Staub:2012pb, Staub:2010jh, Semenov:2008jy, Semenov:2014rea, Alloul:2013bka, Mahmoudi:2008tp,Allanach:2001kg, Djouadi:2002ze, Heinemeyer:1998yj, Porod:2003um, Porod:2011nf, Athron:2014yba, Athron:2017fvs, Eriksson:2009ws, Coimbra:2013qq, Djouadi:1997yw, Muhlleitner:2003vg, Ellwanger:2004xm, Ellwanger:2005dv, Actis:2012qn, Actis:2016mpe, Denner:2017vms, Denner:2017wsf, Das:2011dg}. 
	\item Observable calculators, for example those relating to cross section calculations~\cite{Beenakker:1996ed, Fuks:2013vua, Bonvini:2015ira, Belyaev:2012qa, Kublbeck:1990xc, Hahn:2000kx, Harlander:2012pb, Denner:2014cla, Guzzi:2014wia, Gao:2013kp, Camarda:2019zyx, Grazzini:2017mhc}, 
	muon $g-2$ predictions~\cite{Athron:2015rva, Athron:2021evk, Athron:2022gga}, 
	DM observables~\cite{Backovic:2015cra, Alguero:2023zol, Alguero:2022inz, Belanger:2020gnr, Bringmann:2018lay, Gondolo:2004sc, Bauer:2020jay}, 
	precision Higgs boson measurements~\cite{Bechtle:2008jh, Bechtle:2011sb, Bechtle:2012lvg, Bechtle:2013wla, Bechtle:2015pma, Stal:2013hwa, Bechtle:2014ewa, Bahl:2022igd, Bernon:2015hsa}, 
	neutrino observables~\cite{Wallraff:2014qka, Stowell:2016jfr, IceCube:2012fvn} and cosmological and astronomical observations~\cite{Wainwright:2011kj, Basler:2020nrq, Basler:2024aaf, Athron:2020sbe, Athron:2024xrh, Auffinger:2022dic, Camargo-Molina:2013qva}. 
	\item Collider Monte-Carlo simulation tools~\cite{Alwall:2014hca, Frederix:2018nkq, Sjostrand:2006za, Frixione:2010ra, Frederix:2020trv, Ovyn:2009tx, deFavereau:2013fsa, Conte:2012fm, Denner:2016kdg, Sjostrand:2014zea, Dobbs:2001ck, Buckley:2014ana, Bierlich:2019rhm, Buckley:2010ar, Li:2018qnh, Bellm:2015jjp, Drees:2013wra,Dercks:2016npn, Maguire:2017ypu, Cacciari:2011ma, Gleisberg:2008ta, Kilian:2007gr, Buckley:2021neu, Papucci:2014rja, Guo:2023nfu, Ambrogi:2017neo}. 
	\item Tools using artificial intelligence and machine learning~\cite{Guo:2023nfu, Caron:2016hib, Ren:2017ymm, Goncharov:2021wvd, Wojcik:2023usm, Kriesten:2024are, Bechtle:2013wla, Stal:2013hwa, Bechtle:2014ewa}. 
\end{itemize}
Despite extensive searches, no definitive experimental evidence for BSM has yet been found. 
On the theoretical side, many BSM theories involve complicated structures and large numbers of free parameters, making them difficult to test and constrain systematically. 
Therefore, ``to leave no stone unturned'', there is strong practical demand for tools that can link and execute multiple codes within a single study. 
In this context, several packages have been developed~\cite{Diaz:2024sxg, Staub:2019xhl, Shang:2023gfy, Goodsell:2023iac}. 
Among them, \code{Easyscan\_HEP} focuses on simplifying parameter scans for BSM models.
\code{BSMArt}~\cite{Goodsell:2023iac} provide an all-in-one \code{SARAH} toolbox~\cite{Staub:2013tta, Goodsell:2017pdq, Braathen:2017izn, Goodsell:2018tti, Porod:2003um, Porod:2011nf, Goodsell:2015ira, Staub:2008uz, Alguero:2023zol, Alguero:2022inz, Belanger:2020gnr, Bechtle:2008jh, Bechtle:2011sb, Bechtle:2012lvg, Bechtle:2013wla, Bechtle:2015pma, Bahl:2022igd, Goodsell:2021iwc, Straub:2018kue} that automates the workflow from a model Lagrangian density to the calculation of various observables.  
\par Because BSM parameter spaces are often high-dimensional and the number of experimental constraints continues to grow, assessing the impact of available data on a given model requires a rigorous treatment within a well-defined statistical framework~\cite{AbdusSalam:2020rdj}.
Accordingly, tools for global statistical fits in Bayesian and frequentist frameworks are now well established in the field.
Representative examples include \code{ZFitter} for electroweak precision studies~\cite{Arbuzov:2005ma}, \code{GFitter} for studies of \textit{CP} violation and the CKM matrix~\cite{Baak:2011ze, Charles:2004jd}, \code{NuFIT} for neutrino global fits~\cite{NuFIT, Capozzi:2013csa, Bergstrom:2015rba, Forero:2014bxa}, and \code{SMEFit} for Standard Model effective field theory (SMEFT) analyses~\cite{Ethier:2021bye, Hartland:2019bjb, deBlas:2022ofj}.
For BSM theories, especially in earlier studies of supersymmetric models, widely used tools included \code{MasterCode}\cite{Bagnaschi:2017tru, Bagnaschi:2016afc, Bagnaschi:2016xfg}, \code{SuperBayeS}, \code{SFitter}\cite{Lafaye:2004cn}, and \code{Fittino}\cite{Bechtle:2004pc, Bechtle:2010igv, Bechtle:2012zk}, together with many phenomenological applications and reinterpretations\cite{RuizdeAustri:2006iwb, Bertone:2015tza, Bertone:2011nj, Strege:2012bt, Strege:2011pk, Bridges:2010de, Scott:2009jn, Trotta:2008bp, Baltz:2004aw, Allanach:2005kz, 2020SciPy-NMeth, Balazs:2008ph, Lopez-Fogliani:2009qdp, Meng:2024lmi, Cao:2024axg, Cao:2023gkc, Cao:2022ovk, Cao:2021tuh, Cao:2021lmj, Cao:2019ofo, Cao:2018rix, Cao:2018iyk, Cao:2019evo, Cao:2019qng, Han:2022juu, Yang:2022gvz, Athron:2022uzz, Han:2021gfu, Carragher:2023gcf, Robens:2015gla, ATLAS:2024qmx}.
More recently, \code{GAMBIT}\cite{GAMBIT:2017yxo, GAMBITModelsWorkgroup:2017ilg, GAMBITFlavourWorkgroup:2017dbx, GAMBIT:2017qxg, Martinez:2017lzg, Cornell:2017opo, GAMBITDarkMatterWorkgroup:2017fax, GAMBITCosmologyWorkgroup:2020htv, Bloor:2021gtp}, the ``Global and Modular Beyond the Standard Model Inference Tool'', has provided a broad open-source framework for global fits of BSM theories, combining data from astronomy, cosmology, collider physics, flavour physics, and neutrino physics\cite{GAMBIT:2023yih, Chang:2022jgo, Balazs:2022tjl, Athron:2021auq, GAMBIT:2021rlp, GAMBITCosmologyWorkgroup:2020rmf, Bhom:2020lmk, Kvellestad:2019vxm, Athron:2024rir, Chang:2023cki, Beniwal:2022rde, Chang:2022jgo, Athron:2022afe, AbdusSalam:2020rdj, Rajec:2020orn, Hoof:2018ieb, GAMBIT:2018gjo, GAMBIT:2018eea, Kvellestad:2018akf, Kvellestad:2017rwl, GAMBIT:2017zdo, GAMBIT:2017gge}.
These frameworks play an important role in statistically consistent model testing, but they also reflect the growing complexity of modern phenomenological workflows.

\par Despite the existence of these tools, there remains practical demand for a lightweight framework that can support exploratory studies by individual researchers or small teams with relatively low setup overhead. 
To perform numerical analysis of a physical model, researchers need both a solid grasp of the underlying theory and familiarity with specialised computational tools. 
Early in a research project, mastering both aspects simultaneously can be challenging. 
This creates a practical bottleneck, in which theoretical exploration slows as researchers work through increasingly complex numerical tools.
However, when numerical studies become more accessible in terms of both computational resources and setup requirements, they can accelerate understanding of both the underlying theory and the associated computational methods. 
This paper introduces \code{Jarvis-HEP}, a lightweight Python framework for workflow composition and parameter scans in high-energy physics. It is designed to support parameter scans, external calculations, and result analysis within a lightweight and configurable workflow, while reducing the amount of custom code required from the user.

\par This paper is structured as follows. 
In Sec.~\ref{sec:dsg}, the design philosophy and the main structure of \code{Jarvis-HEP} are briefly summarised. 
Sec.~\ref{sec:yaml} provides a general guide to how users can interact with \code{Jarvis-HEP} through a YAML input file. 
A set of practical examples is given in Sec.~\ref{sec:scan} to provide an intuitive comparison of different sampling methods. 
Sec.~\ref{sec:example} presents an end-to-end \code{EggBox} example showing how to deploy scan tasks, monitor execution, inspect generated outputs, and understand the resulting data. 
A summary and suggested future directions are given in Sec.~\ref{sec:sum}.

\section{Design Philosophy and Key Features}
\label{sec:dsg}
\begin{figure*}[th]
	\centering
	\includegraphics[width=0.98\linewidth]{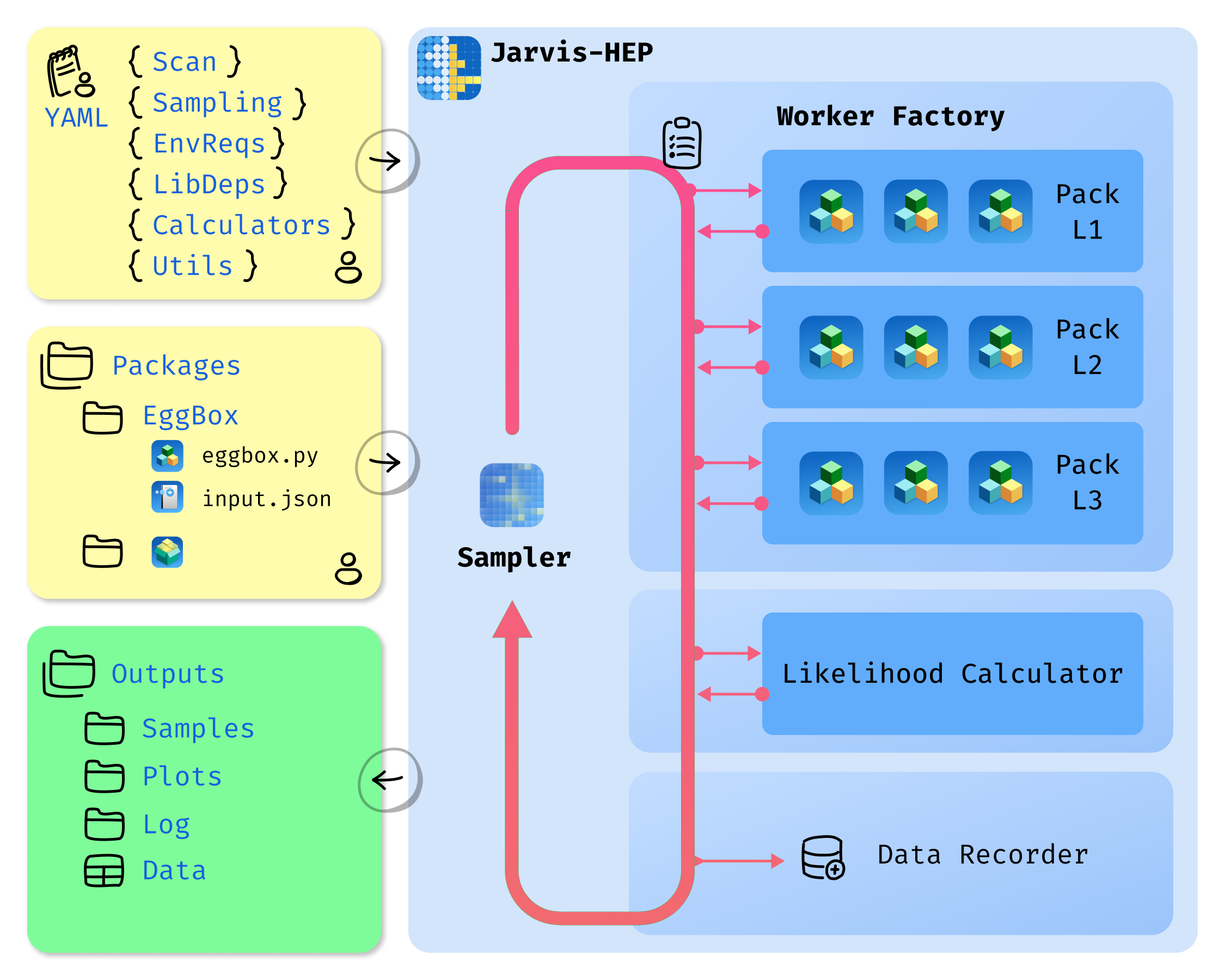}
    \caption{\label{fig:workflow}Conceptual overview of the 
    \code{Jarvis-HEP} workflow. The YAML input file and packages in the {\bf yellow boxes} are user-provided. Items in the green box are outputs from \code{Jarvis-HEP}. The blue box illustrates the architecture and workflow of \code{Jarvis-HEP}. The "Worker Factory" manages multiple computing layers (\code{L1}, \code{L2}, and \code{L3}), each containing one or more computing module pools, which handle a computing software package (\code{Pack}). A task submitted by the sampler is executed through the calculation flow (red circle). The \code{Likelihood Calculator} evaluates observables, and results are recorded in the \code{Data Recorder}. Arrows in the red circle indicate asynchronous communication between the sampler, worker factory, and computation modules.}
\end{figure*}
The \code{Jarvis-HEP} framework streamlines high-energy physics (HEP) model analysis by leveraging modularity, parallelism, and automation for efficiency and scalability. Typically, HEP workflows involve intricate data pipelines, numerous software dependencies, and computationally demanding tasks, such as parameter space sampling, observable calculations across various packages (as detailed in Sec.~\ref{sec:int}), and likelihood evaluations. \code{Jarvis-HEP} tackles these challenges with a structured yet adaptable framework that incorporates asynchronous task scheduling, automated dependency management, and concurrent execution of physics calculations. Figure~\ref{fig:workflow} illustrates the \code{Jarvis-HEP} workflow, showing the progression of tasks through the system:
\begin{itemize}
	\item The user must provide a YAML input file (see \href{https://www.yaml.org}{www.yaml.org}) that specifies how to perform a model scan and calculate observables or likelihoods. 
  Detailed instructions are available in Sec.~\ref{sec:yaml}.
	\item In Fig.~\ref{fig:workflow}, \code{Packages} refers to the source code files of the packages invoked by the Worker factory (\code{Calculators} in the YAML input file) or their dependencies (\code{LibDeps} in the input \code{YAML}). 
	\item In \code{Jarvis-HEP}, a task represents a sample that corresponds to a parameter point in the BSM model. Each task is assigned a unique identifier (UUID) from the start, when the sampler generates a random number array. 
	\item The worker factory is a singleton object responsible for managing global task scheduling. When The YAML input file is parsed, \code{Jarvis-HEP} automatically processes the input and output dependencies of all calling packages, generating a multi-layer workflow in the form of a directed hierarchical flow graph. Each computing package is depicted as a module node. For each package, the factory generates a module pool with multiple replicas. These replicas are managed to ensure that only one task is processed at a time, avoiding conflicts in paths and intermediate calculations. Packages with independent inputs and outputs are grouped within the same computation layer to minimize overall task time. The factory depicted in Fig.~\ref{fig:workflow} comprises three computing layers (\code{L1}, \code{L2}, and \code{L3}), each containing three computing packages. When a task is submitted, the factory distributes it across the layers simultaneously. After completing calculations within a layer, the factory consolidates the information from all computing packages before passing it to the next layer. Once all calculations are finished, the factory calls the \code{Likelihood Calculator} to compute the likelihood. It then simultaneously submits the task information to the \code{Data Recorder} and returns it to the sampler. 
	\item In Section~\ref{sec:yaml}, we will introduce the \code{Likelihood Calculator}, as defined in the YAML input file.
	\item The \code{Data Recorder} is a standalone process that receives data in real-time and writes it to an HDF5 file at regular intervals, with a default interval of 30 seconds.
	\item The \code{Sampler} generates random number sequences using the chosen algorithm. The supported sampling algorithms will be detailed in Sec.~\ref{sec:scan}. In \code{Jarvis-HEP}, all sampling algorithms are implemented as iterators. 	
\end{itemize}
\par The standout feature of \code{Jarvis-HEP} is its asynchronous parallel processing. Unlike traditional synchronous execution, which handles tasks sequentially, asynchronous execution allows multiple computations to run simultaneously. This non-blocking approach minimizes delays and improves system responsiveness. Faster nodes can process more tasks independently of slower ones, leading to more balanced and efficient workloads. This efficiency also applies to I/O operations and program logging.
\par Modular design is a key feature of \code{Jarvis-HEP}, allowing the framework to adapt to various research needs. By following the abstract factory pattern, it offers a structured approach to adding new features while keeping the architecture clean and organized. This design simplifies the integration of sampling algorithms, enabling the incorporation of different numerical methods without changing the framework's core structure. The modularity also extends to input and output (IO) methods, supporting multiple file formats. Researchers can integrate \code{Jarvis-HEP} with external high-energy physics tools using custom IO handlers, ensuring compatibility across different computing environments. This modular approach allows for the independent addition of user-friendly features such as configuration management, error handling, and logging, without affecting existing functionality. As a result, both novice and experienced users can enjoy a flexible and efficient research workflow.
\par Thanks to its pure Python implementation, \code{Jarvis-HEP} is easy to install. It only requires a specific Python version and a few libraries. A brief installation and quick start guide can be found in \ref{app:ins}.

\section{User interface}
\label{sec:yaml}

In this section, we describe the user interface of \code{Jarvis-HEP}. The framework is operated through a command-line interface and configured via YAML input files. Tasks are organised in self-contained project directories, where workflows, external programs, and data processing steps are specified declaratively. \code{Jarvis-HEP} then handles execution, dependency management, and structured output automatically.

We first summarise the command-line usage, then introduce the project-based workflow, and finally present the structure of the YAML configuration file.

\subsection{Project-based workflow}
\label{sec:project}

\code{Jarvis-HEP} organises computational tasks using a project-based workflow, in which each project directory represents a self-contained computational unit. A project can be created via
\begin{lstterm}
Jarvis project create MyProject
\end{lstterm}
which initializes a structured workspace for defining and running scan tasks.

A typical project directory contains subdirectories such as \code{bin/} (YAML configuration files and entry points), \code{data/} (input datasets), and \code{deps/} (external programs or resources). The project root is identified by configuration files such as \code{.jarvis-project.json} and \code{jarvis.project.yaml}. During execution, \code{Jarvis-HEP} automatically generates directories including \code{outputs/}, \code{logs/}, \code{images/}, and \code{checkpoints/} to store results, logs, visualizations, and checkpoint states.

This project-based design provides a unified structure for managing configurations, dependencies, and outputs, ensuring reproducibility, portability, and efficient reuse in practical research workflows. Additional project management commands and options can be accessed via
\begin{lstterm}
Jarvis project -h
\end{lstterm}
and are documented in the official \code{Jarvis} documentation.

\subsection{Command-line interface}

The \code{Jarvis-HEP} framework is operated through the command-line tool \code{Jarvis}. A typical workflow is initiated by providing a YAML configuration file that defines the scan task:
\begin{lstterm}
Jarvis myinput.yaml
\end{lstterm}
This command parses the configuration, constructs the workflow, and executes the scan within the current project environment.

The available command-line options can be listed via:
\begin{lstterm}
Jarvis -h
\end{lstterm}
For completeness, we summarise the main options below:
\begin{description}[style=nextline, leftmargin=1cm]
	\item[\code{-h, -{}-help}] Display usage information and available options.
	\item[\code{-d, -{}-debug}] Run \code{Jarvis} in debug mode, increasing the verbosity of log output. This option is typically used in conjunction with \code{-{}-check-modules}.
	\item[\code{-{}-check-modules}] Test the workflow by sampling a small number of parameter points to validate module connections.
	\item[\code{-{}-skip-library-installation}] The packages described in \code{LibDeps} in \file{myinput.yaml} will skip the installation operation. This usually happens after these packages have already been installed.
	\item[\code{-{}-skip-draw-flowchart}] Skip generation of the workflow flowchart to reduce overhead.
	\item[\code{-{}-convert}] Convert the dataset from \code{hdf5} format to \code{csv} format for analysis. This can be done in a separate terminal while the scanning task is running, without compromising the safety of the main scanning process.
	\item[\code{-{}-monitor}] Monitor the current Jarvis-HEP job in real-time to track CPU, memory, and I/O usage. This should be done in a separate terminal while the scan task is running.
\item[\code{-{}-plot}] Generate a \code{Jarvis-PLOT} YAML configuration file with default plotting settings, which can then be used as the starting point for visualizing scan results\footnote{Installation and usage of \code{Jarvis-PLOT} are documented in the unified online \code{Jarvis} documentation portal, which covers all components of the \code{Jarvis} ecosystem.}.
\end{description}

 \subsection{YAML input file}
 \label{sec:yaml:scan}
YAML (Yet Another Markup Language) is a human-readable data-serialization format that provides a structured way to define workflows in \code{Jarvis-HEP}. In the framework, the YAML input file specifies the complete configuration of a scan task, including sampling strategies, environment requirements, external dependencies, computational modules, and auxiliary utilities. As illustrated in Fig.~\ref{fig:workflow}, the top-level structure of the YAML file is organised into six sections:
\begin{description}
	\item [\ykey{Scan}] defines general information about the scan task.
	\item [\ykey{Sampling}] specifies the sampling algorithm, parameter definitions, and likelihood evaluation.
	\item [\ykey{EnvReqs}] describes the environment requirements for executing the workflow.
	\item [\ykey{LibDeps}] defines external dependencies required by computational modules.
	\item [\ykey{Calculators}] specifies the computational modules used to evaluate observables.
	\item [\ykey{Utils}] defines interpolation functions for use in symbolic expressions. 
\end{description}
The structure is illustrated using a representative example YAML configuration.
\subsubsection{\ykey{Scan} section}
\begin{lstyaml}
Scan: 
  name:     "EggBox_Random_01"
  save_dir: "&J/outputs"
\end{lstyaml}
The \ykey{Scan} section provides a simple example of YAML syntax, in which colons separate \ykey{keys} and \yvalue{values}.
\begin{itemize}
	\item \ykey{name}: a string value that specifies the name of the scan task.
	\item \ykey{save\_dir}: is a string that specifies the path where all outputs of this scan task are saved. The symbol \yvalue{\&J} represents the root directory of the current \code{Jarvis} project. Additional special address symbols are listed in \ref{app:addsyb}.
\end{itemize}
\subsubsection{\ykey{Sampling} section}
\label{yaml:sampling}
\begin{lstyaml}
Sampling:
  Method: "Random" 
  Variables:
    - name: x
      description: "log scale distributed var x"
      distribution:
        type: Log
        parameters:
          min: 0.1
          max: 10.0

    - name: y
      description: "flat distributed var y"
      distribution:
        type: Flat
        parameters:
          min: 0
          max: 5.0
  
  Point number: 1000

  LogLikelihood: 
    - {name: "LogL_Z", expression: "LogGauss(z, 100, 10)"}
  selection: "log(2.0 * x) < sin(y)"
\end{lstyaml}
The \ykey{Sampling} section outlines the sampling procedure. It is essential to complete the \ykey{Variables}, \ykey{LogLikelihood}, and \ykey{selection} settings, as they are common and must not be left blank.
\begin{itemize}
\item \ykey{Method}: This specifies the sampling method as a string value. Here, it is set to "Random," meaning points are sampled randomly. Additional sampling methods will be introduced in Section~\ref{sec:scan}.
\item \ykey{Variables}: This sector lists the sampling variables, each represented as a dictionary containing the following keys:
\begin{itemize}
\item \ykey{name}: The variable name as a string. All variables can be used in the symbolic expression, see \ref{app:expr}. 
\item \ykey{description}: A string that briefly explains the variable.
\item \ykey{distribution}: Defines the variable's probability distribution. A summary of all distributions and their properties can be found in \ref{app:dist}.
\end{itemize}
\item \ykey{Point number}: The total number of sampling points, expressed as an integer. 
\item \ykey{LogLikelihood}: This section lists the log-likelihood functions, each defined by:
\begin{itemize}
\item \ykey{name}: The function name as a string. 
\item \ykey{expression}: A symbolic mathematical expression written in Python syntax. In this context, it refers to $\log{\mathcal{L}_z} = -\frac{(z - 100)^2}{2 \times 10^2}$, where $z$ is the output defined in Section \ref{sec:calc}. Additional symbolic expressions and examples can be found in Appendix \ref{app:expr}.
\end{itemize}
\item \ykey{selection}: A logical expression, see \ref{app:expr}. Filters sampled points based on a specified condition. Selected points will be sent to the Worker Factory for observable calculation. Samples not meeting the selection condition will be discarded.
\end{itemize}
\subsubsection{\ykey{EnvReqs} section}
\begin{lstyaml}
EnvReqs:
  OS: 
    - name: linux 
      version: ">=3.10.0"
    - name: Darwin
      version: ">=10.0"
  Check_default_dependencies:
    required: True 
    default_yaml_path:  "&J/deps/environment_default.yaml"	
\end{lstyaml}
The \ykey{EnvReqs} section outlines the environmental requirements necessary for running the \code{Jarvis-HEP} framework effectively. Users should configure their systems to meet these conditions prior to execution.
\begin{itemize}
\item \ykey{OS}: a list specifying the supported operating systems along with their required versions. Each entry contains:
\begin{itemize}
\item \ykey{name}: a string representing the OS name. Both "linux" and "Darwin" (macOS) are included.
\item \ykey{version}: a string indicating the minimum OS version required.
\end{itemize}
\item \ykey{Check\_default\_dependencies}: a dictionary indicating whether to verify default dependencies prior to execution.
\begin{itemize}
\item \ykey{required}: A boolean value indicating if the dependency check is mandatory. If set to \yvalue{True}, the system will enforce the check before execution.
\item \ykey{default\_yaml\_path}: This text specifies the path to the default dependency configuration file. The placeholder ``\yvalue{\&J}'' represents the root directory of the current \code{Jarvis} project. For more details on directory resolution, refer to \ref{app:addsyb}. To streamline system settings for multiple similar tasks, users can store their frequently used system requirements in a separate YAML file and reference its path here.
\end{itemize}
\end{itemize}
\subsubsection{\ykey{LibDeps} section}
\label{sec:libs}
The \code{EggBox} calculator in this example does not require the \ykey{LibDeps} packages. In this section, we demonstrate the functionality of the \ykey{LibDeps} section by installing two high-energy physics packages.
\begin{lstyaml}[caption={Example of \ykey{LibDeps} Configuration section}]
LibDeps:
  path: "&J/External/Library"
  make_parallel: 16
  Modules:
  - name: "HepMC"
    required_modules: []
    installed: False
    installation:
      path: "&J/External/Library/HepMC"
      source: "&J/External/Library/Source/HepMC-2.06.09.tar.gz"
      commands:
      - "cd ${LibDeps:path}"
      - "rm -rf HepMC*"
      - "cp ${source} ./"
      - "tar -xzf HepMC-2.06.09.tar.gz"
      - "cd ${LibDeps:path}/HepMC-2.06.09"
      - "./bootstrap"
      - "./configure --with-momentum=GEV --with-length=MM --prefix=${path}"
      - "make -j${LibDeps:make_parallel}"
      - "make install"

  - name: "Pythia8"
    required_modules: 
    - "HepMC"
    installed: False
    installation:
      path: "&J/External/Library/Pythia8"
      source: "&J/External/Library/Source/pythia8230.tgz"
      commands:
      - "cd ${LibDeps:path}"
      - "mkdir Pythia8"
      - "cd ${LibDeps:path}/Source"
      - "tar -zxvf pythia8230.tgz"
      - "cd ${LibDeps:path}/Source/pythia8230"
      - "./configure --with-hepmc2=${LibDeps:path}/HepMC --prefix=${path}"
      - "make -j${LibDeps:make_parallel}"
      - "make install"	$
\end{lstyaml}

The \ykey{LibDeps} section handles external dependencies by specifying installation paths, parallel compilation settings, and necessary libraries.
\begin{itemize}
\item \ykey{path}:  Specifies the root directory for external libraries.
\item \ykey{make\_parallel}: Sets the number of parallel compilation jobs to optimize build time.
\item \ykey{Modules}: The list of libraries, each with its own installation details, includes two high-energy physics software packages: \code{HepMC} and \code{Pythia8}. For illustration, this example installs both packages, noting that \code{Pythia8} depends on the prior installation of \code{HepMC}.
\end{itemize}

Each module contains:
\begin{itemize}
\item \ykey{name}: Library identifier.
\item \ykey{required\_modules}: Dependencies that must be installed beforehand.
\item \ykey{installed}: A Boolean flag indicating whether the library is already installed. If true, the package installation will be skipped. 
\item \ykey{installation}:
\begin{itemize}
\item \ykey{path}: Installation destination directory.
\item \ykey{source}: Source tarball location.
\item \ykey{commands}: A sequence of shell commands to automate extraction, configuration, compilation, and installation.
\end{itemize}
\end{itemize}

Here we demonstrate the complete installation process of two high-energy physics packages, including their decompression, compilation, and installation. The commands provided are exactly what users would enter in the terminal, ensuring a smooth and efficient transition of existing workflows to \code{Jarvis-HEP}. \yvalue{HepMC} has no dependencies and is configured using \yvalue{GEV} and \yvalue{MM} units. In contrast, \yvalue{Pythia8} relies on \yvalue{HepMC} and requires its installation path for configuration. 
\par In \code{Jarvis-HEP}, the syntax \yvalue{\$\{\ykey{keys}\}} is used for self-referencing (see Section \ref{sec:selfref}). For instance, the compilation command \yvalue{make -j\$\{\ykey{LibDeps}:\ykey{make\_parallel}\}} enables parallel processing. Here, the value \yvalue{16} from the \ykey{make\_parallel} key in the \ykey{LibDeps} section is substituted into \yvalue{\$\{LibDeps:make\_parallel\}}, resulting in the command \yvalue{make -j16} after parsing.

\subsubsection{\ykey{Calculators} section}
\label{sec:calc}
\begin{lstyaml}
Calculators:
  make_parallel: 16
  path: "&J/WorkShop/Program"
  Modules:
  - name: EggBox
    required_modules: [] 
    clone_shadow: True
    path: &eggbox_path "&J/WorkShop/Program/EggBox/@PackID"  # Define Anchor point 
    source: "&J/External/Inertial/EggBox"
    installation:
    - "cp -r ${source}/* ${path}"
    initialization:
    - "cp -r ${source}/input.json ${path}/input.json"
    - "rm -f output.json"
    execution:
      path: *eggbox_path  # using the Anchor point
      commands:
      - "./eggbox.py"	
      input:
      - name: inpjson
        path: "&J/WorkShop/Program/EggBox/@PackID/input.json"  
        type: "JSON"
        actions:
        - type: "Dump"
          variables:
          - {name:"xx", expression:"x*Pi"}
          - {name:"yy", expression:"y*Pi"}
          save: True
      output:
      - name: oupjson
        path: "&J/WorkShop/Program/EggBox/@PackID/output.json" 
        type: "JSON"
        save: True
        variables:
        - {name: z}
\end{lstyaml}
\begin{figure*}[th]
	\includegraphics[width=0.98\linewidth]{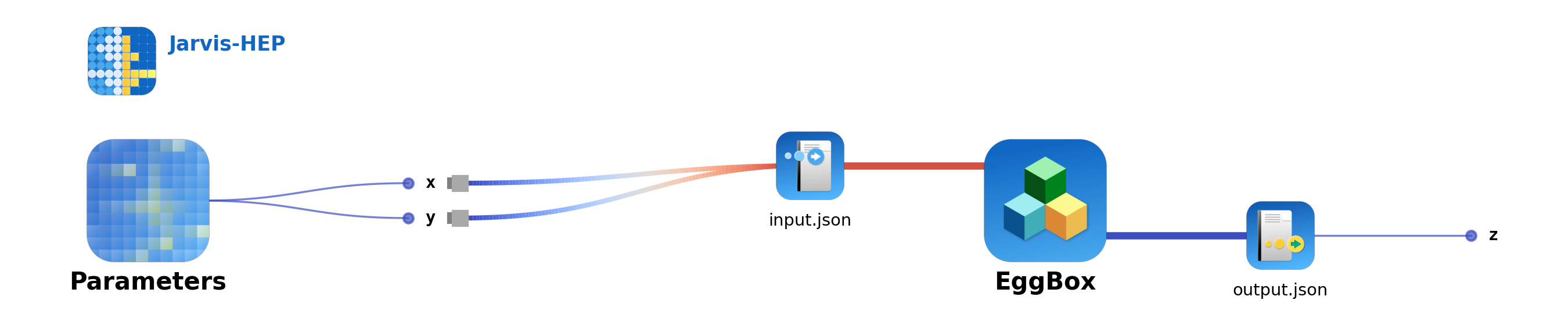}
	\caption{\label{fig:fceggbox}Flow chart of scan defined in \file{bin/Example\_Random.yaml}, which is automatically generated by \code{Jarvis-HEP}.}
\end{figure*}
The \ykey{Calculators} section defines the computational setup for the module \code{EggBox}. The YAML snippet above details the computational setup, execution workflow, and I/O handling for the specified module, \code{EggBox}. It includes the following elements:
\begin{itemize}
    \item \ykey{make\_parallel}: Specifies the level of parallel execution to 16. At the same time, it also sets the number of replicas for the calculators. 
    \item \ykey{path}: the factory's base directory.
    \item The \ykey{Modules} entry is similar to the \ykey{Modules} entry in \ykey{LibDeps}, but  it requires additional settings.
\end{itemize}
The \ykey{Modules} section here lists a specific computational module to be executed: a Python script called \code{EggBox}.
\begin{itemize}
    \item \ykey{name}: The module's name.
    \item \ykey{required\_modules}: This setting is similar to the one in \ykey{LibDeps}. It is a list that includes all the dependent packages specified in \ykey{LibDeps} and \ykey{Calculators}. \code{EggBox} does not rely on any additional modules or packages.
	\item \ykey{clone\_shadow}: Set to \yvalue{True} to enable the module to create and execute replicas.
    \item \ykey{path}: The directory is defined using the YAML anchor \yvalue{\&eggbox\_path}, which represents the path \yvalue{\&J/WorkShop/Program/EggBox/@PackID}. Here, \yvalue{\&J} refers to the current project root, and \yvalue{@PackID} is a placeholder that can be used in directory-like values. For more details, refer to \ref{app:addsyb}.  
    \item \ykey{source}: Refers to the external source directory of \code{EggBox}.
    \item \ykey{installation}: Similar to \ykey{LibDeps} module, a list of installation commands is specified here. 
    \item \ykey{initialization}: To prevent program corruption from multiple calls during runtime, users should initialize the package before calculations by clearing previous outputs and restoring the input template file.
    \item \ykey{execution}: The detailed execution steps are as follows: The \ykey{calculator} module operates as a black box, receiving \ykey{input} from one or more files. It performs calculations using terminal \ykey{commands} within a specified \ykey{path} and produces \ykey{output} to a file. The I/O handling is described below.
\end{itemize}

In the workflow shown in Fig.~\ref{fig:fceggbox}, \code{Jarvis-HEP} manages data exchange between different components through structured input and output files. The framework supports multiple file formats, such as \code{SLHA} and \code{JSON}, to ensure compatibility with a wide range of HEP tools, and this support will continue to be extended. In this example, we use the \code{JSON} format as a representative case.

In the \ykey{input} block of the \ykey{execution} section above, the input file \yvalue{inpjson} is read from the specified \code{JSON} path. During processing, a \yvalue{Dump} action is applied to define derived variables, where \ykey{xx} and \ykey{yy} are computed as \yvalue{x * Pi} and \yvalue{y * Pi}, respectively. The transformed values are then written to the input file used by the calculator. In the \ykey{output} block, the output file \yvalue{oupjson} is read from the specified location in \code{JSON} format. The variable \ykey{z}, produced by the computation, is extracted and saved as part of the structured scan output.

As illustrated in Fig.~\ref{fig:fceggbox}, \code{EggBox} is a computational unit designed to calculate a function.
\begin{equation}
	\code{z} = \left( \sin(\code{x}\pi) \cos(\code{y}\pi) + 2 \right)^5. 
\end{equation}
The input file and the output file of \code{EggBox} are both \code{JSON} type files. Input file \file{input.json} is given as follows,  
\begin{lstyaml}
{ 
	"xx":	10.0,
	"yy":	10.0
}
\end{lstyaml}
where input variables are defined using mathematical expressions, 
\begin{equation}\label{eq:eggbox}
	\code{xx} = \code{x}  \pi, \quad \code{yy} = \code{y} \pi. 
\end{equation} 
The output file for the above input is given as follows.  
\begin{lstyaml}
{ 
	"z":	89.44573638583269, 
	"Time":	0.18490937203932312
}
\end{lstyaml}
As shown above, the scan parameters \code{x} and \code{y} are not used directly; instead, they are used in mathematical expressions. \code{Jarvis-HEP} can automatically parse symbols, parameters, input and output variables, mathematical constants, and function operations. For more details, see \ref{app:expr}. When the worker factory is created, a logic flow chart, like the one in Fig.~\ref{fig:fceggbox}, is automatically generated. This greatly simplifies the user's workflow and enhances deployment efficiency. All commands that need execution adhere to specific parsing rules (see Section \ref{sec:selfref}). Besides the \code{JSON} format shown here, \code{Jarvis-HEP} also supports other formats, such as \code{SLHA}. We will keep adding support for more formats. Please refer to the \href{https://github.com/Pengxuan-Zhu-Phys/Jarvis-HEP}{online documentation} for details.

\subsubsection{\ykey{Utils} section}
\label{sec:utils}
\begin{lstyaml}
Utils:
  interpolations_1D:
  - name: XenonSD2019
    file: "&J/External/Info/Xenon1T2019SD_p.csv"
    logY: True
    logX: False 
    kind: "cubic"
  - name: inter1P
    x_values: [1, 2, 3, 4, 5]
    y_values: [2, 1.8, 1.5, 1.2, 0.8]
    kind: "cubic"	
\end{lstyaml}
In addition to the built-in functions listed in \ref{app:expr}, \code{Jarvis-HEP}'s symbolic expression parsing function allows users to define custom functions using data. These custom functions can be used just like the built-in ones. Here, we present two methods for declaring one-dimensional interpolation functions, as listed in \ykey{interpolations\_1D}. The first method involves creating an interpolation function based on the spin-dependent dark matter-nucleon scattering cross-section upper limit data from the Xenon experiment released in 2019. 
\begin{itemize}
	\item \ykey{name}: The function name is included in the expression. In this example, you can call the function using \yvalue{XenonSD2019(50.)} to input the value 50.
	\item \ykey{file}: The directory containing the \code{CSV} file with interpolation data. The data file must include only two original columns, labeled as \yvalue{x,y}: 
		\begin{lstyaml}
x,y
6.06E+00,1.15E-37
6.15E+00,9.57E-38
...	
		\end{lstyaml}
	\item \ykey{logX}, \ykey{logY}: These Boolean values determine whether to use a logarithmic scale for interpolation. If not specified, the default value is \yvalue{False}.
	\item \ykey{kind}: Specify the type of interpolation either as a string or an integer indicating the order of the spline interpolator to use. The string should be one of the following: \yvalue{zero}, \yvalue{slinear}, \yvalue{quadratic}, or \yvalue{cubic}, corresponding to spline interpolation of zeroth, first, second, or third order, respectively.
\end{itemize}
The second method involves providing data directly using two lists: \ykey{x\_values} and \ykey{y\_values}. Ensure both lists are of the same length. The \ykey{file} option is not needed, and the remaining options are the same as in the first method. 
\par In addition to defining interpolation functions directly in this \ykey{Utils} section, users can also provide them through the companion project \code{Jarvis-Operas}\footnote{\code{Jarvis-Operas} is available from PyPI: \url{https://pypi.org/project/Jarvis-Operas/}. In \code{Jarvis-Operas}, interpolation and other reusable numerical functions are implemented and registered, then loaded into a \code{Jarvis-HEP} workflow for use in symbolic expressions.
This is useful when the same interpolation functions are shared across multiple scan projects, or when users prefer to maintain these functions in a separate Python package rather than directly in the YAML file.}. 

\subsubsection{Three self-reference syntaxes in Jarvis YAML}
\label{sec:selfref}
Writing a YAML input file for a scan task involves three additional features beyond plain text.
\begin{itemize}
	\item Anchors and aliases. 
	\par YAML supports anchors and aliases for reusing values within a document. An anchor is defined using \yvalue{\&}, and later referenced using \yvalue{*}. For example:
	\begin{lstyaml}
- &flag Apple
- Blueberry 
- *flag
	\end{lstyaml}
	After parsing, the alias is replaced by the anchored value:
	\begin{lstyaml}
- Apple
- Blueberry 
- Apple
	\end{lstyaml}
	An anchor can be referenced multiple times, and aliases always refer to the most recently defined anchor.
	
	\item Short self-referencing in command. 
	\par Short self-referencing uses the \yvalue{\$\{key\}} notation to reference values within the same \ykey{module}, applicable to both \ykey{LibDeps} and \ykey{Calculators}. For instance, consider the installation command \yvalue{"cp \$\{source\} ./"} for the HepMC package, as shown in \ref{sec:libs}. When this command is parsed, the value of \ykey{source} for the HepMC module is substituted, resulting in the execution of the command \yvalue{"cp \&J/External/Library/Source/\ HepMC-2.06.09.tar.gz ./"}.

	\item Long self-referencing in command. 
	\par Long self-references allow access to values outside the current module in a YAML structure. They are used to reference global or higher-level keys. The correct syntax involves starting from the top-level section key and separating each level's keys with colons until reaching the desired level. For example, \yvalue{\$\{LibDeps:make\_parallel\}} refers to the number 16, and \yvalue{\$\{LibDeps:path\}} points to the root directory for the external library \yvalue{"\&J/External/Library"}.
\end{itemize}
Please note that long and short self-referencing can only be used in commands. Specifically, this applies to the \ykey{installation} in \ykey{LibDeps} modules, and the \ykey{installation}, \ykey{initialization}, and \ykey{commands} in \ykey{execution} options within \ykey{Calculators} modules.

\section{Sampling methods in \code{Jarvis-HEP}}
\label{sec:scan}
As illustrated in Fig.~\ref{fig:workflow}, the sampling components in \code{Jarvis-HEP} are architecturally decoupled from the worker factory, thereby facilitating the seamless integration of diverse sampling algorithms within the framework.
The samplers are implemented as Python iterators, providing a flexible and memory-efficient paradigm for parameter space exploration.
Through the iterator design pattern, \code{Jarvis-HEP} enables lazy evaluation of sampling processes,
wherein parameter values are generated on-demand rather than pre-computed and stored in memory.
This approach confers significant advantages for high-dimensional parameter spaces by eliminating unnecessary memory overhead, ensuring computational scalability,
and enhancing both performance and usability.

\subsection{Background on sampling algorithms}
\label{sec:sampling-background}

Before introducing the sampler interfaces implemented in \code{Jarvis-HEP}, we briefly review several basic ideas used in parameter-space exploration. The purpose of this subsection is to provide the minimal statistical background needed to understand the behaviour of the supported samplers, while the following subsections focus on their concrete YAML configuration and usage within the framework.

\subsubsection{Inverse transform sampling}
The inverse transform sampling algorithm~\cite{devroye2006nonuniform} is used to sample from given distributions with an analytical functional form,
such as prior distributions with a known closed form for the Cumulative Density Function (CDF).
The essential idea of the algorithm is to transform samples from a
uniform distribution with support $[0, 1]$ by the inverse of the CDF function $F(\cdot)$ of the target distribution $f(\cdot)$.
Specifically, we can get a sample $\mathbf{x}$ from the target distribution by following the transformation:
\begin{align}
    F^{-1}: \omega \mapsto \mathbf{x}, 
\end{align}
where $\omega$ denotes the sample from the uniform distribution.
To explain the inverse sampling method in more detail, consider the example of a truncated Jeffreys prior~\citep{gnedenko2018theory} for a scale parameter which follows an exponential distribution. The Probability Density Function (PDF) of the truncated Jeffreys prior is 
\begin{align}
    p(\theta) = \frac{1}{\theta\ln{(U/L)}}, \quad \theta\in[L, U].
\end{align}
We obtain the CDF and its inverse as
\begin{align}
    &F(\theta) = \frac{\ln{(\theta/L)}}{\ln{(U/L)}}, \\
    &F^{-1}(u) = L\cdot(U/L)^{u}.
\end{align}
With the above inverse of CDF function $F^{-1}$,
we can sample from the Jeffery truncated prior distribution by first sampling from the uniform distribution with support $[0, 1]$,
and then computing $\theta=F^{-1}(u)$.

\subsubsection{Importance sampling}
In Bayesian analyses, the posterior distribution is written as
\begin{equation}
P(\bm{\theta}|\mathcal{D}) = \frac{P(\mathcal{D}|\bm{\theta})P(\bm{\theta})}{P(\mathcal{D})},
\end{equation}
where \(\bm{\theta}\) denotes the model parameters and \(\mathcal{D}\) denotes the observed data. 
Importance sampling estimates posterior quantities by drawing samples from a proposal distribution \(q(\bm{\theta})\) and assigning weights proportional to
\begin{equation}
w(\bm{\theta}) = \frac{P(\mathcal{D}|\bm{\theta})P(\bm{\theta})}{q(\bm{\theta})}.
\end{equation}
A simple choice is to use the prior distribution as the proposal, \(q(\bm{\theta}) = P(\bm{\theta})\), in which case the unnormalised weight reduces to the likelihood,
\begin{equation}
w(\bm{\theta}) = P(\mathcal{D}|\bm{\theta}).
\end{equation}
This provides a likelihood-weighted prior sample and can be useful for simple exploratory studies, reweighting, and evidence estimates in low-dimensional or well-matched problems.

\par The main limitation of this approach is weight degeneracy. 
If the likelihood is sharply peaked compared with the prior volume, most prior samples receive negligible weights, and the posterior estimate is dominated by a small number of points. 
This makes direct prior-proposal importance sampling inefficient in high-dimensional parameter spaces, but the underlying idea of likelihood weighting remains useful in more advanced sampling methods and in the interpretation of randomly generated scan datasets.

This idea can also be illustrated with a deterministic grid. 
Suppose the parameter space is discretised into grid points \(\{\bm{\theta}_i\}\). 
If these points are assigned weights
\begin{equation}
w_i \propto \mathcal{L}(\mathcal{D}|\bm{\theta}_i)\pi(\bm{\theta}_i),
\end{equation}
then the resulting weighted grid provides a discrete representation of the likelihood-weighted parameter space. 
For uniform priors and equal grid-cell volumes, the weights are proportional to the likelihood values at the grid points. 
This interpretation is useful for visualisation and low-dimensional exploratory studies, although the number of grid points grows rapidly with dimension.

\par The main limitation of simple prior-based or grid-based weighting is weight degeneracy. 
If the likelihood is sharply peaked compared with the prior volume, most sampled or gridded points receive negligible weights, and the posterior estimate is dominated by a small number of points. 
This is one reason why more adaptive methods, such as Markov-chain and nested-sampling approaches, are often needed for high-dimensional phenomenological studies.

\subsection{Independent and deterministic samplers}
In the current version, \code{Jarvis-HEP} supports several simple sampling strategies that do not rely on Markov-chain evolution, including random sampling, grid sampling, and Bridson sampling. 
These samplers generate parameter points without constructing complex mechanisms like Markov chains, and are useful for exploratory scans, validation tests, and visualising the behaviour of the scan workflow.

\begin{figure*}[t]
	\centering
	\makebox[\textwidth][c]{	
	\hspace{-0.4cm}	
	\begin{subfigure}[t]{0.4\textwidth}
		\includegraphics[width=\textwidth]{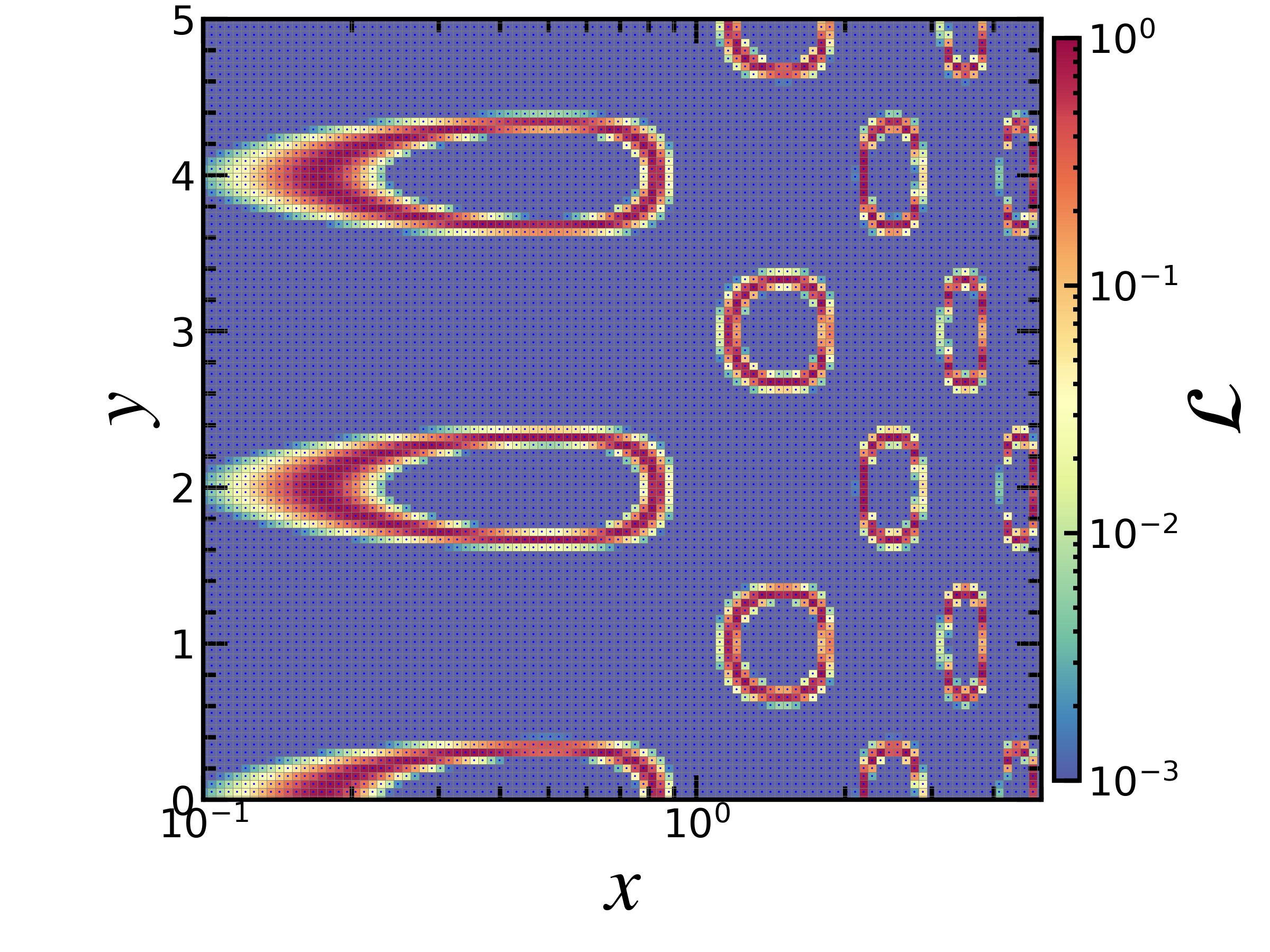}
		\caption{Grid sampling}
	\end{subfigure}
	\hspace{-0.3cm}
	\begin{subfigure}[t]{0.4\textwidth}
		\includegraphics[width=\textwidth]{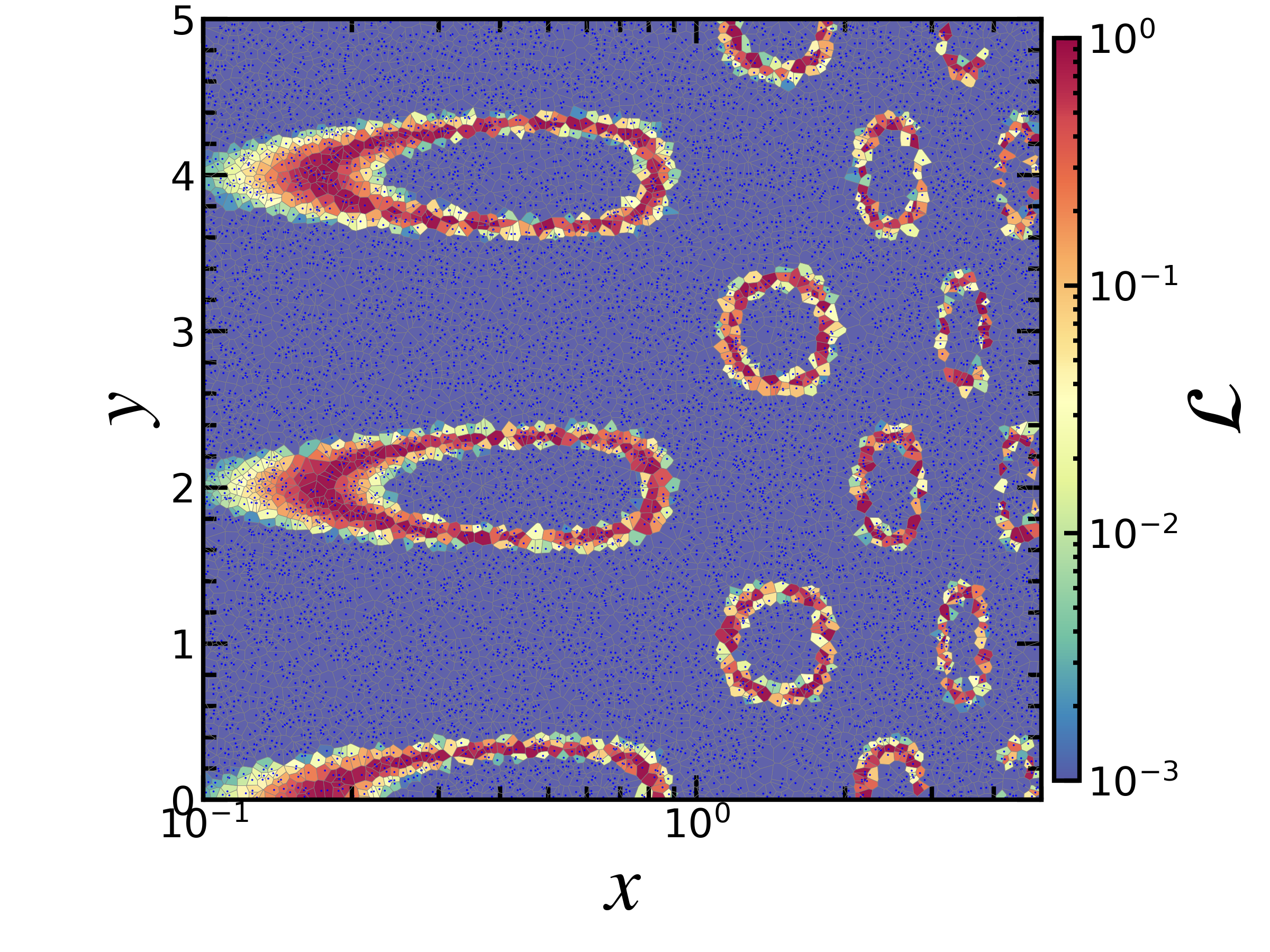}
		\caption{Random sampling}
	\end{subfigure}
	\hspace{-0.3cm}
	\begin{subfigure}[t]{0.4\textwidth}
		\includegraphics[width=\textwidth]{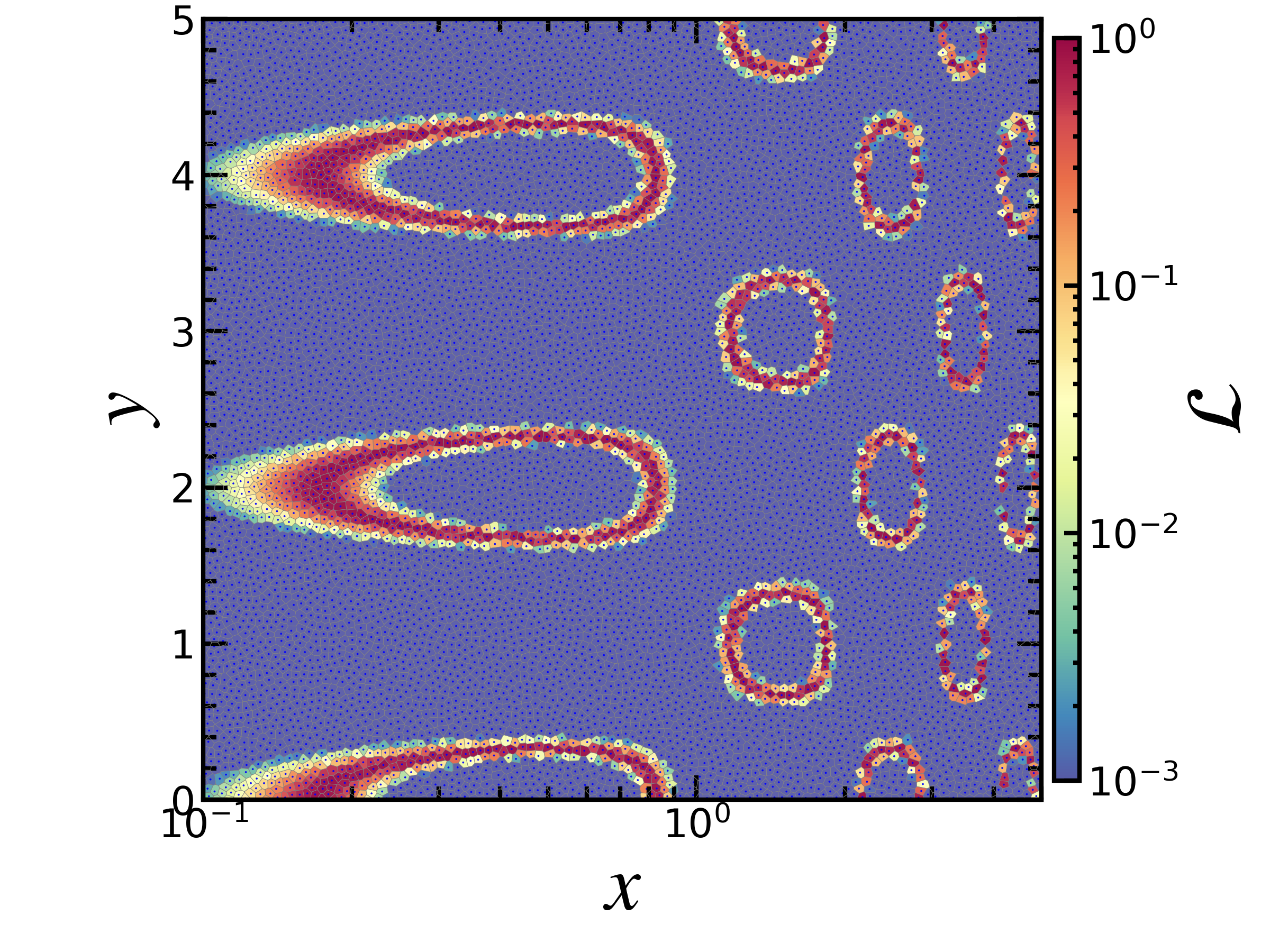}
		\caption{Bridson sampling}
	\end{subfigure}
	}
	\caption{\label{fig:simsam}Data visualisation using Voronoi diagram for grid, random, and Bridson sampling results for the \code{EggBox} model. Each plot contains 10,000 samples. Cell color coded with likelihood value $\mathcal{L}$. }
\end{figure*}

\subsubsection{Random sampling}
The \yvalue{Random} sampler draws independent parameter points from the user-defined prior distributions. 
For each sampled point, \code{Jarvis-HEP} applies the selection rule, executes the calculator workflow for accepted points, evaluates the requested likelihood terms, and records the result. 
The resulting dataset may be interpreted as a likelihood-weighted prior sample, which is closely related to the simple importance-sampling construction described above. 
In the default \code{Jarvis-HEP} workflow, however, the purpose of the \yvalue{Random} sampler is to generate and record evaluated parameter points rather than to perform a full importance-sampling inference procedure.

A random scan can be configured as follows:
\begin{lstyaml}
Sampling:
  Method: "Random"
  Variables: ...
  Point number: 1000
  LogLikelihood:
  - {name: "LogL_Z",  expression: "LogGauss(z, 100, 10)"}
  selection: ...
\end{lstyaml}

Here, \ykey{Method} selects the random sampler, while \ykey{Point number} specifies the number of parameter points to be generated. 
The \ykey{Variables} block defines the prior distributions, \ykey{selection} filters the generated points before calculator execution, and \ykey{LogLikelihood} specifies the likelihood terms to be evaluated and recorded.

\subsubsection{Grid sampling}
The \yvalue{Grid} sampler provides a deterministic scan over a discretised parameter space. 
Compared with random sampling, it gives a regular coverage of the selected parameter ranges and is therefore useful for low-dimensional exploratory studies, validation tests, and visual comparisons of sampling behaviour. 
Since the number of grid points grows rapidly with the number of scan variables, this method is mainly intended for scans with only a few dimensions.

In \code{Jarvis-HEP}, grid sampling can be configured as follows:
\begin{lstyaml}
Sampling: 
  Method: "Grid"
  Variables: 
  - name: x
    description: "Variable X"
    distribution:
      type: Flat
      parameters: {min: 0, max: 5.0, num: 5}

  - name: y
    description: "Variable Y"
    distribution:
      type: Flat
      parameters: {min: 0, max: 5.0, num: 5}
  
  LogLikelihood: ...
  selection: ... 
\end{lstyaml}

In this example, \ykey{Method} selects the grid sampler. 
For each variable, \ykey{min} and \ykey{max} define the scanned range, while \ykey{num} specifies the number of evenly spaced grid points in that direction. 
The total number of generated points is the product of the \ykey{num} values over all scanned variables. 
The \ykey{selection} and \ykey{LogLikelihood} entries are handled in the same way as in other sampling methods.

\subsubsection{Bridson sampling}
For HEP models whose parameter spaces exhibit structured patterns-specifically,
where high-probability regions are well-separated by distances greater than $r$,
it is desirable to incorporate this prior information into the posterior sampling procedure.
One effective approach is to employ a prior that enforces such spatial separation among parameter samples.
Following this principle, we adopt the Bridson algorithm (also known as the blue noise sampler) to generate parameter samples that reflect this repulsive spatial structure~\cite{bridson2007fast}~\footnote{An elegant and intuitive visualization of the Bridson algorithm is available in \href{https://bost.ocks.org/mike/algorithms/}{Mike Bostock's illustrated note}.}.
The Bridson algorithm, a variant of Poisson disk sampling,
generates samples that are uniformly distributed with a minimum separation distance,
thereby avoiding clustering while maintaining randomness.
This property makes it particularly suitable for encoding repulsive priors in parameter spaces where diversity among samples is desirable.  It allows for a more even coverage of the space, outperforming truly random sampling~\cite{bergstra2012random}.
\begin{lstyaml}
Sampling:
  Method: "Bridson" 
  Variables:
  - name: X
    description: "Variable X"
    distribution:
      type: Flat
      parameters: {min: 0, max: 100, length: 100}

  - name: Y
    description: "Variable Y"
    distribution:
      type: Log
      parameters: {min: 100, max: 150, length: 50}

  Radius: 6
  MaxAttempt:  30
  LogLikelihood: ...
  selection: ...	
\end{lstyaml}
The example above performs Bridson sampling in the two-dimensional \yvalue{x}-\yvalue{y} plane.
Specifically, it maps the parameter ranges $\yvalue{x} \in [0, 100]$ and $\yvalue{y} \in [100, 150]$ 
onto a rectangular domain of width \yvalue{100} in the \yvalue{x} direction and height \yvalue{50} in the \yvalue{y} direction.
A minimum inter-sample distance of \yvalue{6}, specified via the \ykey{Radius} parameter, ensures a repulsive sampling pattern,
while the rejection tolerance is controlled by setting \ykey{MaxAttempt} to \yvalue{30}.
Although the original Bridson algorithm supports sampling in arbitrary dimensions,
our empirical analysis indicates that in high-dimensional settings,
the auxiliary grid structure used for spatial acceleration incurs substantial memory overhead.
To mitigate this, \code{Jarvis-HEP} restricts Bridson sampling to at most five dimensions.

\par To compare the performance of the three sampling strategies: grid sampling,
uniform random sampling, and Bridson sampling,
we conduct a two-dimensional parameter scan over the space: \begin{equation} x \in [0.1, 5], \quad y \in [0, 5], \end{equation}
where $x$ is sampled logarithmically and $y$ linearly.
The resulting sample distributions, rendered using the plotting utilities of \code{Jarvis-HEP}, are shown in Fig.~\ref{fig:simsam}.
In Fig.~\ref{fig:simsam}, we visualize the sample patterns generated by each method.
Grid sampling (left) yields a regular lattice structure, while uniform random sampling (center) produces an irregular distribution with varying Voronoi cell sizes, reflecting its stochastic nature.
In contrast, Bridson sampling (right) enforces a minimum separation between points,
resulting in a more uniform yet non-grid-like pattern,
characterized by relatively balanced Voronoi regions.

\subsection{Markov Chain Monte Carlo}
\begin{figure*}[th]
	\centering
	\makebox[\textwidth][c]{	

	\hspace{-0.4cm}
	\begin{subfigure}{0.4\textwidth}
	\includegraphics[width=\textwidth]{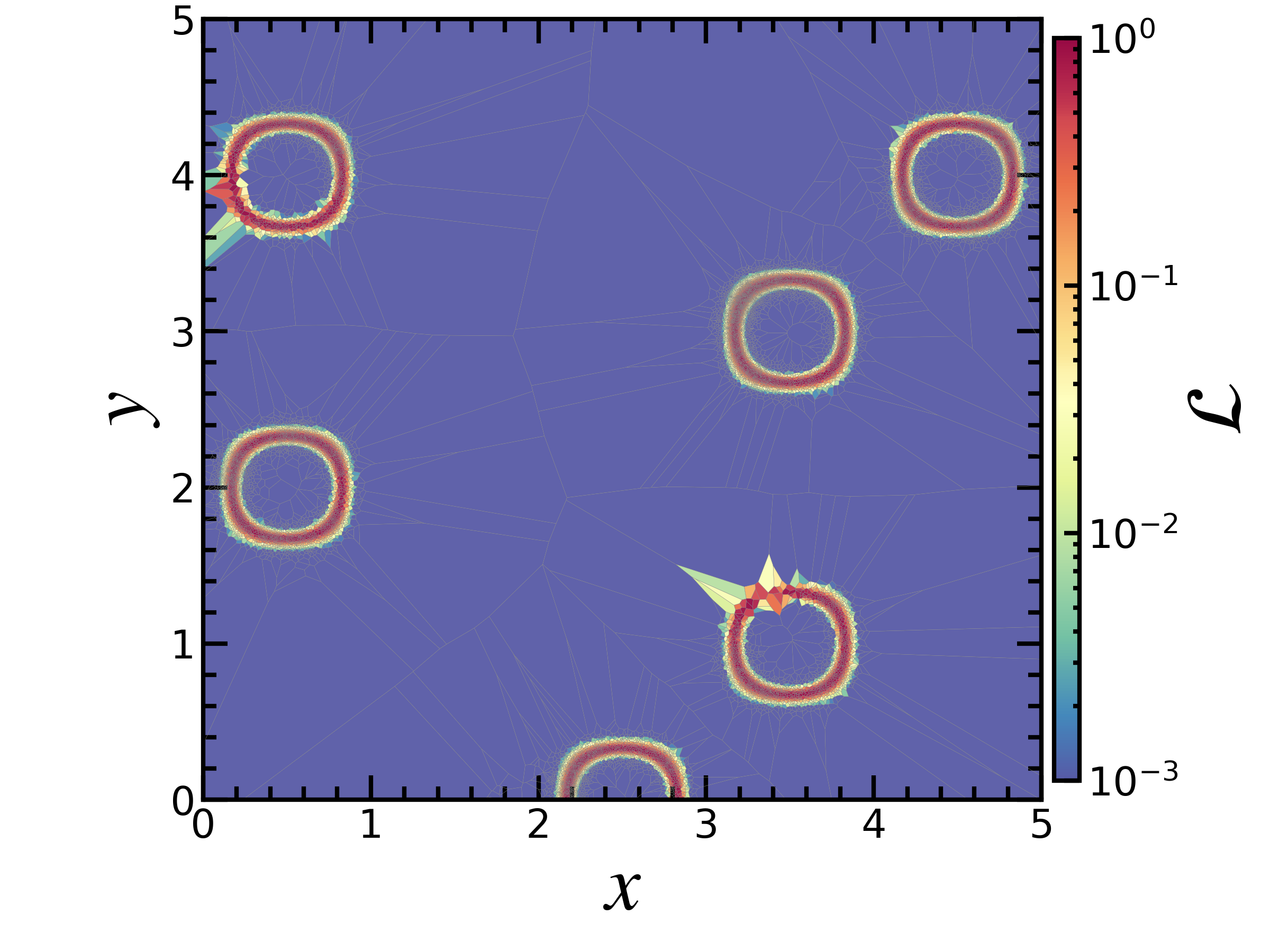}
	\caption{\ykey{proposal\_scale} = 0.01}
        \label{fig:small_prop_std}
	\end{subfigure}
	\hspace{-0.3cm}
	\begin{subfigure}{0.4\textwidth}
	\includegraphics[width=\textwidth]{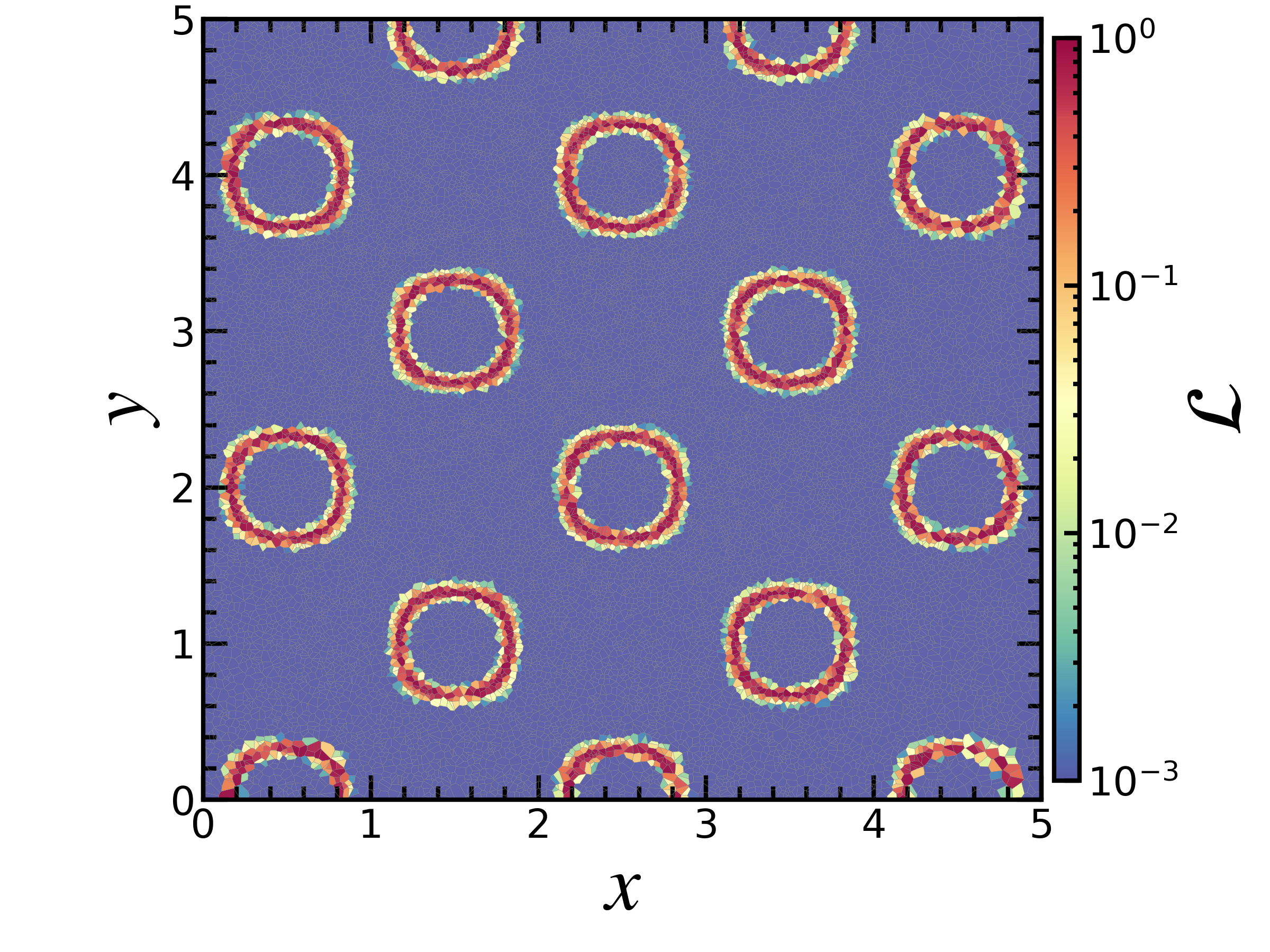}
	\caption{\ykey{proposal\_scale} = 0.20}
        \label{fig:big_prop_std}
	\end{subfigure}
	\hspace{-0.3cm}
	\begin{subfigure}{0.4\textwidth}
	\includegraphics[width=\textwidth]{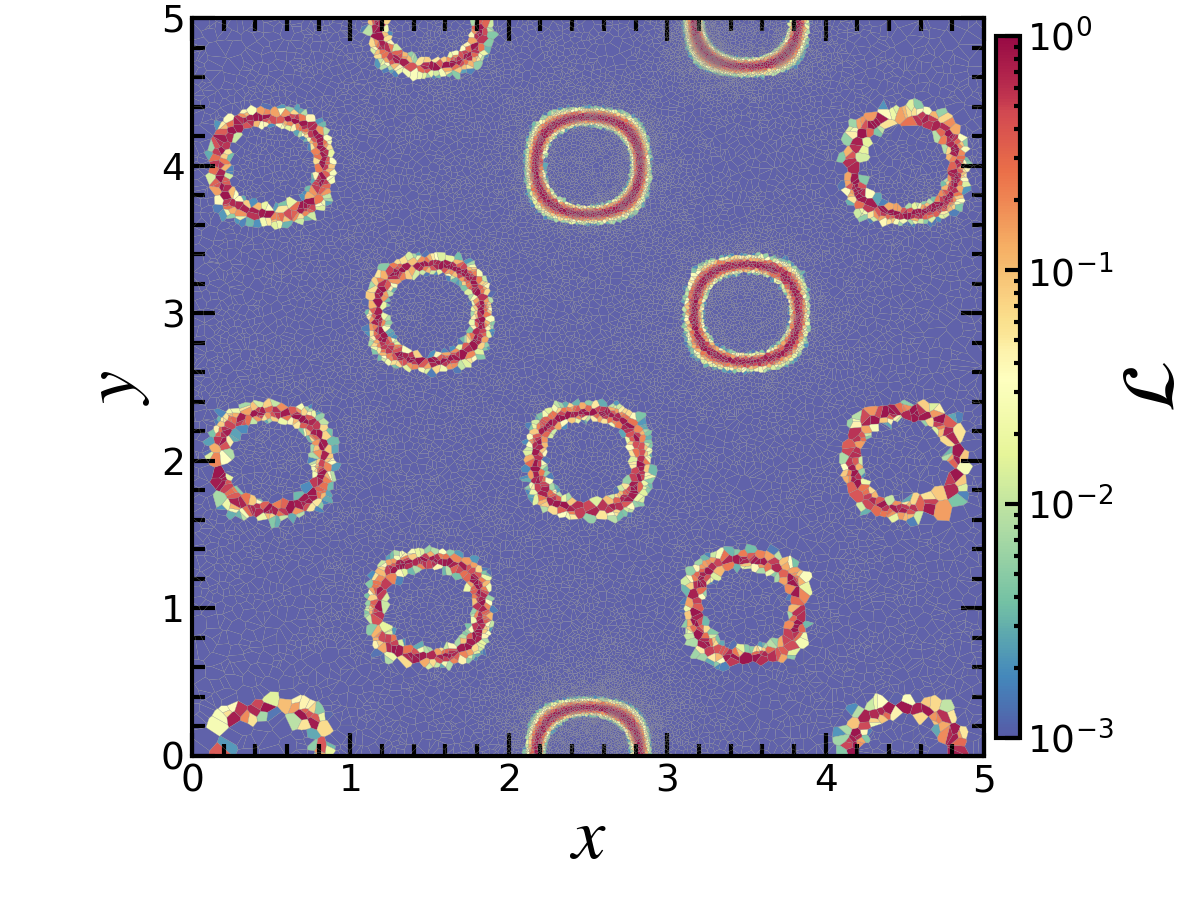}
	\caption{PT-MCMC sampling}
	\end{subfigure}
	}
	\caption{\label{fig:mcmc} Similar to Fig.~\ref{fig:simsam}, but using MCMC sampling methods. Panels (a) and (b) show standard MCMC scans with different proposal scales, illustrating the trade-off between local resolution and global exploration: (a) small proposal steps and (b) large proposal steps. Panel (c) shows the PT-MCMC result, in which multiple chains with different proposal scales perform exchange steps. Each panel contains 25,000 points generated using 10 chains.}
\end{figure*}

In Bayesian data analysis, Markov Chain Monte Carlo (MCMC) algorithms are fundamental and powerful tools,
which provide straightforward approaches to Bayesian statistical inference.
Due to the flexibility of the MCMC framework, as well as its ease of implementation,
MCMC techniques are widely used in almost all areas, including phenomenological analysis in HEP.
\code{Jarvis-HEP}, currently supports the Metropolis-Hastings (MH) algorithm and
the parallel tempering MCMC algorithm (PT-MCMC).

\subsubsection{Metropolis-Hastings method}
The fundamental principle underlying MCMC is the construction of a Markov Chain
whose stationary distribution converges to the target posterior distribution $P(\boldsymbol{\theta}|\mathbf{D})$.
The theoretical foundation of MCMC relies on the detailed balance condition,
which ensures that the chain converges to the correct stationary distribution.
For a Markov chain with transition probability $T(\boldsymbol{\theta}' \leftarrow \boldsymbol{\theta})$, detailed balance requires:
\begin{align}
    \pi(\boldsymbol{\theta})T(\boldsymbol{\theta}' \leftarrow \boldsymbol{\theta}) = \pi(\boldsymbol{\theta}^{\prime})T(\boldsymbol{\theta} \leftarrow \boldsymbol{\theta}^{\prime})
\end{align}
where $\pi(\boldsymbol{\theta})$ represents the target distribution.
This condition, combined with reversibility of the chain, guarantees that the equilibrium distribution is indeed the desired posterior.
The ergodicity property ensures that the chain can reach any state with non-zero probability, enabling comprehensive exploration of the parameter space.
The MH algorithm exemplifies these principles through its acceptance-rejection mechanism.
Given a current state $\boldsymbol{\theta}$ and a proposed state $\boldsymbol{\theta}'$ drawn from a proposal distribution $q(\boldsymbol{\theta}'| \boldsymbol{\theta})$,
the acceptance probability is:
\begin{align}
    \text{Accept}(\boldsymbol{\theta}', \boldsymbol{\theta}) = \min{\bigg(1, \frac{\pi(\boldsymbol{\theta}')q(\boldsymbol{\theta}|\boldsymbol{\theta}')}{\pi(\boldsymbol{\theta})q(\boldsymbol{\theta}'|\boldsymbol{\theta})}\bigg)}
\end{align}
For the common case of symmetric proposals where $q(\boldsymbol{\theta}' | \boldsymbol{\theta}) = q(\boldsymbol{\theta}|\boldsymbol{\theta}')$,
this simplifies to the ratio of posterior probabilities, making the algorithm particularly straightforward to implement for phenomenological studies where only relative probabilities are required.
Convergence assessment is crucial for ensuring reliable posterior inference.
The convergence of MCMC chains to the stationary distribution is typically evaluated using several diagnostic metrics.
The Gelman-Rubin statistic $\hat{R}$ compares within-chain and between-chain variances across multiple independent chains, with values close to unity indicating convergence~\citep{gelman1992inference}.
The effective sample size $N_{\text{eff}}$ quantifies the number of independent samples obtained from the correlated chain, accounting for autocorrelation effects.
Additionally, trace plots and autocorrelation functions provide visual diagnostics for assessing mixing efficiency and burn-in periods,
ensuring that the sampled chain has adequately explored the posterior distribution before statistical inference is performed.

\subsubsection{MCMC sampler}
\par The MH sampling method in \code{Jarvis-HEP} can be conducted via the following YAML settings: 
\begin{lstyaml}
Sampling:
  Method: "MCMC" 
  Variables:
  - name: X
    description: "Variable X"
    distribution:
      type: Flat
      parameters: {min: 0, max: 5}
  - name: Y
    description: "Variable Y"
    distribution:
      type: Flat
      parameters: {min: 0, max: 5}
  Bounds: 
    num_chains: 10
    num_iters: 2500
    proposal_scales: 0.1
  LogLikelihood: ...
  selection: ...
\end{lstyaml} 
Three parameters need to be specified in \ykey{Bounds}:
\begin{itemize}
	\item \ykey{num\_chains}: an integer which declares the number of independent MCMC chains. 
	\item \ykey{num\_iters}: the length of each chain. 
	\item \ykey{proposal\_scales}: $Q(y_{i+1}|y_i)$ is a high-dimensional Gaussian function which acts in uniform space $x=F(y)$. The peak position is at $x_i = F(y_i)$, and the width is a number in the range $[0, 1]$. 
\end{itemize}
Fig.~\ref{fig:mcmc} illustrates the effects of \ykey{proposal\_scale} on the performance of the MH algorithm.
In Fig.~\ref{fig:big_prop_std}, we apply a small $\ykey{proposal\_scale}=0.01$ in the simulation.
It can be observed that the samples efficiently capture the fine details around local likelihood maxima.
However, if the \ykey{proposal\_scale} is too small it severely restricts global exploration, causing chains to get trapped in isolated regions and failing to explore the parameter space adequately.
Conversely, utilizing a larger $\ykey{proposal\_scale} = 0.2$ in MH can substantially improve global coverage, as shown in Fig.~\ref{fig:big_prop_std}, allowing chains to traverse multiple peaks efficiently.
Nevertheless, this results in a cost of reduced local resolution, as the samples may frequently overshoot narrow, high-likelihood regions.
Obtaining optimum performance with MH samplers thus requires carefully tuning the \ykey{proposal\_scale} to balance global and local sampling.
 
\subsubsection{Parallel tempering MCMC algorithm}
The parallel tempering MCMC algorithm (PT-MCMC) is an advanced sampling technique designed to improve the exploration of complex, multimodal probability distributions.
It extends the standard MH algorithm by running multiple Markov chains in parallel at different inverse temperatures $\beta_{i} = \frac{1}{T_{i}}$,
where higher-temperature chains facilitate exploration by traversing energy barriers more freely while lower-temperature chains ensure convergence to the target distribution.
Periodically, adjacent chains propose state swaps, which are accepted with probability
\begin{align}
    P_{\text{swap}} = \min(1, e^{(\beta_{i}E(x_{i}) - \beta_{j}E(x_{j}))}),
\end{align}
where $E(\cdot)$ denotes the energy function.
Specifically, the above formula represents the acceptance probability for Markov chain $i$ to transition to a proposed state $x_{j}$ proposed by an adjacent chain $j$.  
This mechanism allows information exchange between chains, enhancing mixing and improving convergence. PT-MCMC is particularly effective in statistical physics, Bayesian inference, and machine learning, where conventional MCMC methods struggle with slow mixing due to rugged energy landscapes and isolated modes \citep{geyer1991markov, earl2005parallel}.

Within the present context of high-energy physics model analysis, a refinement of the simulation scheme's configuration is necessary. 
Specifically, we consider a scenario where the inverse temperature parameter is set to $\beta=1$, effectively removing its explicit dependence.
The energy function for each Markov chain $i$ is parameterized by a distinct set of hyperparameters, denoted as $\theta_{i}$,
such that the energy of a state $x_{i}$, is given by $E(x_{i}, \theta_{i})$.
Consequently, the reduced acceptance probability for a proposed state swap between adjacent chains $i$ and $j$ is expressed as:
\begin{align}
    P_{\text{swap}} = \min(1, e^{(E(x_{i},\, \theta_{i}) - E(x_{j},\, \theta_{j}))}).
\end{align}

\begin{lstyaml}
Sampling:
  Method: "PTMCMC" 
  Variables:
  - name: x
    description: "Variable X"
    distribution: ... 
  - name: y
    description: "Variable Y"
    distribution: ... 
  Bounds: 
    num_chains: 10
    num_iters:  2500
    exchange_interval: 20
    proposal_scales: [0.3, 0.25, 0.2, 0.15, 0.1, 0.05, 0.03, 0.01, 0.005, 0.002]
  LogLikelihood: ...
  selection: ...
\end{lstyaml}
The above settings describe a scan of the \code{EggBox} model using PT-MCMC with the following options
\begin{itemize}
	\item \ykey{num\_chains}: an integer which sets the number of independent MCMC chains. 
	\item \ykey{num\_iters}: sets the length of each chain. 
	\item \ykey{exchange\_interval}: an integer used to specify how many samples each Markov chain collects before performing an exchange step.
	\item \ykey{proposal\_scales}: a list that contains the proposal widths of different chains.  
\end{itemize}
The scan results are shown in Fig.~\ref{fig:mcmc} [c], demonstrating the effectiveness of the PT-MCMC sampling algorithm. Compared to the simple MCMC method shown in Fig.~\ref{fig:mcmc},
the PT-MCMC approach substantially improves global exploration efficiency, successfully identifying and sampling from multiple distinct high-likelihood regions.
The sampling points in Fig.~\ref{fig:mcmc} [c] show a more uniform and balanced coverage of the parameter space, capturing all significant peaks without becoming trapped in isolated modes.
This enhanced performance results from the strategy of utilizing multiple chains at different \yvalue{proposal\_scales}, defined by the standard deviation of their respective proposal distributions,
allowing exchanges between chains to escape local likelihood maxima more effectively.
Consequently, PT-MCMC achieves a superior balance between local accuracy and global exploration given the same number of samples as the simple MCMC. Therefore, for a complex model, users can increase the length of the chain to explore the parameter space in more detail.

\subsection{Nested sampling method}
\begin{figure}[th]
	\centering\hspace{-0.3cm}
	\includegraphics[width=0.48\textwidth]{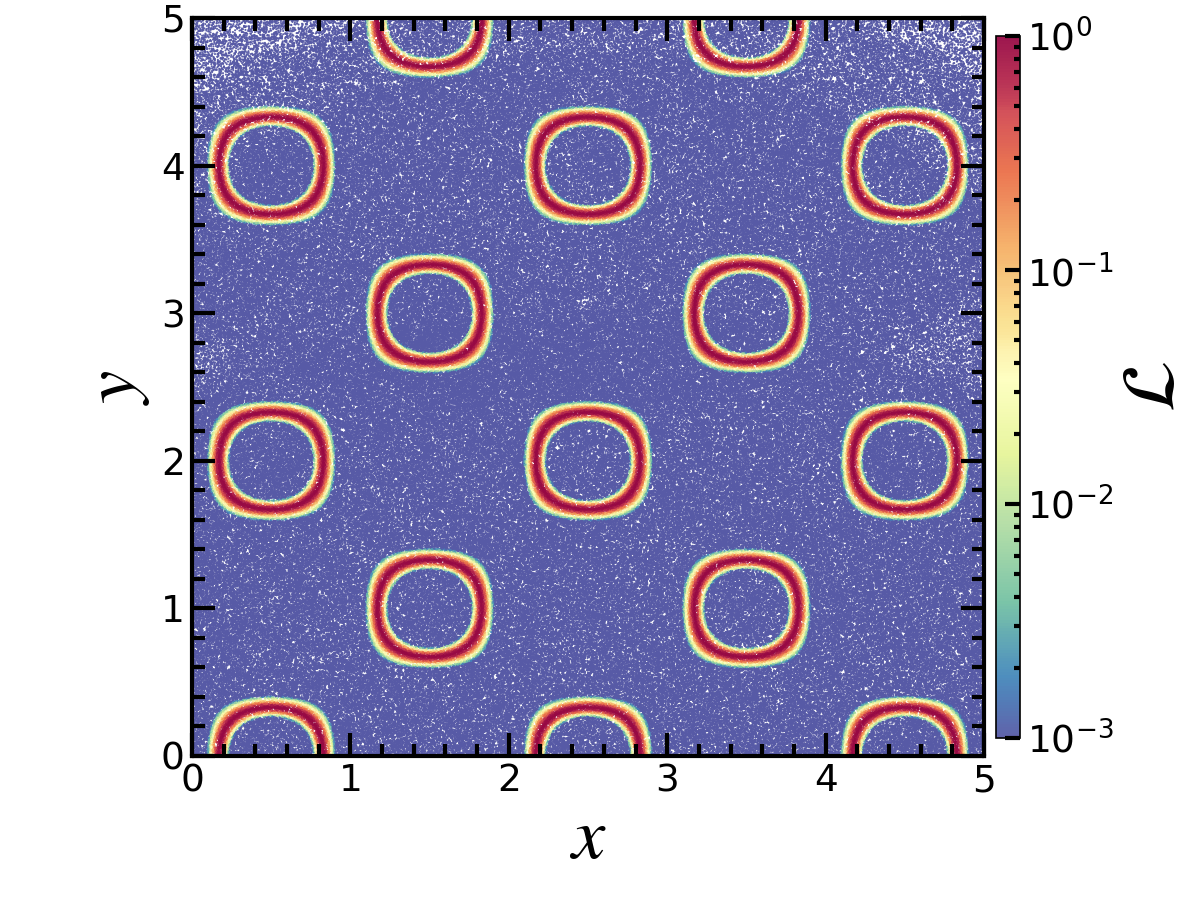}
	\caption{\label{fig:dynt} Scatter plot of \code{EggBox} model using the \code{dynesty} implementation of the nested sampling method. Color coded with likelihood. }
\end{figure}
Nested sampling represents a sophisticated Monte Carlo technique that simultaneously computes the Bayesian evidence (marginal likelihood)
and generates posterior samples through systematic exploration of nested iso-likelihood contours.
Developed initially in Refs.~\citep{skilling2004nested, skilling2006nested}, the method transforms the multidimensional integration problem of evidence calculation
into a one-dimensional problem by exploiting the relationship between prior volume and likelihood values.
The fundamental insight of nested sampling lies in the transformation of the evidence integral:
\begin{equation}
\mathcal{Z} = \int{\rm d}\boldsymbol{\theta}~ \mathcal{L}(\boldsymbol{\theta
    }) \pi(\boldsymbol{\theta
    }) 
\end{equation}
into a one-dimensional integral over the prior volume $X$.
By defining the prior volume enclosed by the iso-likelihood surface $\mathcal{L}(\boldsymbol{\theta}) > L$ as:
\begin{align}
\label{eq:priorV}
X(L) = \int_{\mathcal{L}(\boldsymbol{\theta}) > L}\pi\boldsymbol{\theta})d\boldsymbol{\theta} 
\end{align}
the evidence can be rewritten as:
\begin{equation}
\mathcal{Z} = \int_{0}^{1}\mathcal{L}(X)dX 
\end{equation}
where $L(X)$ represents the likelihood as a function of prior volume.
\begin{figure}[th]
	\centering
	\includegraphics[width=0.9\linewidth]{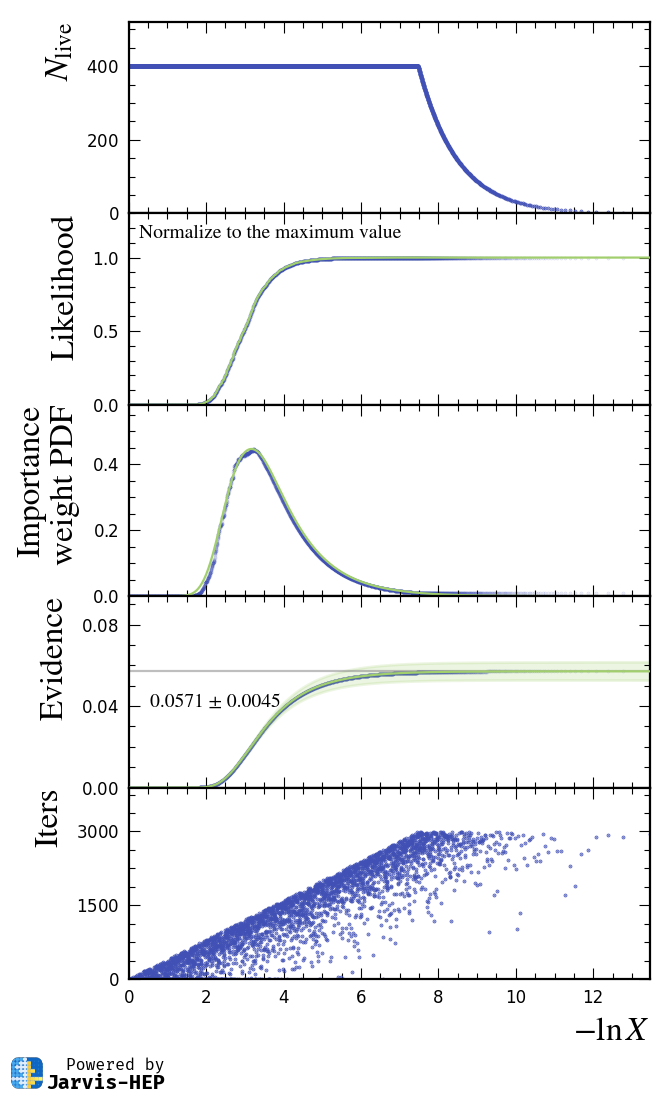}
	\caption{\label{fig:dynesty} The behaviour of \code{dynesty} algorithm in \code{EggBox} model sampling task. See also Fig.~2 in Ref.~\cite{Higson_2018} for a detailed explanation. }
\end{figure}
This transformation enables systematic shrinkage of the prior volume through ordered likelihood sampling.
The nested sampling algorithm maintains a collection of $N$ ``live points'' $\{\boldsymbol{\theta}^{(j)}\}_{j=1}^N$ sampled from the prior distribution.
At each iteration $i$, the algorithm identifies the point with minimum likelihood $L_i = \min_j \mathcal{L}(\boldsymbol{\theta}^{(j)})$
and estimates the prior volume contraction as $X_i = t_i X_{i-1}$,
where $t_i \sim \text{Beta}(N,1)$ follows from order statistics theory.
The discarded point contributes to the evidence estimate:
\begin{align}
\Delta{\mathcal{Z}}_{i} = L_{i}(X_{i-1} - X_{i}).
\end{align}
The critical step involves replacing the discarded point with a new sample drawn from the constrained prior $\pi(\boldsymbol{\theta} | \mathcal{L}(\boldsymbol{\theta}) > L_i)$.
This is achieved through MCMC sampling initialized from one of the remaining live points,
accepting only steps that maintain the likelihood constraint.
This constrained sampling ensures statistical consistency while enabling efficient exploration of progressively smaller high-likelihood regions.
The algorithm terminates when the remaining prior volume becomes negligible,
typically when $X_i < \epsilon$ for some predetermined tolerance.
The final evidence estimate is obtained as $\mathcal{Z} = \sum_i \Delta \mathcal{Z}_{i} + X_{\text{final}} \bar{L}_{\text{final}}$,
where $\bar{L}_{\text{final}}$ represents the mean likelihood of the final live points.

The nested sampling algorithm systematically explores the likelihood surface by maintaining a fixed ensemble of live points and progressively removing
those with lowest likelihood while introducing replacements sampled from the constrained prior above the current threshold.
This process effectuates a controlled contraction of the prior volume, systematically ascending through likelihood levels while maintaining statistical rigor in evidence computation.
As iterations advance, the live points naturally fragment into clusters concentrating around posterior maxima,
preserving sampling efficiency and rendering the algorithm well-suited for multimodal likelihood structures commonly encountered in phenomenological analyses.
However, nested sampling exhibits susceptibility to the curse of dimensionality,
as the efficiency of the constrained MCMC sampling deteriorates exponentially with increasing parameter space dimensionality.
This challenge is exacerbated by the geometric concentration of probability mass
in narrow posterior regions characteristic of high-dimensional spaces,
requiring increasingly sophisticated sampling strategies to maintain algorithmic reliability.

\par For the \code{EggBox} example, deployment of a \code{dynesty} implemention of nested sampling can be configured with the following \code{YAML}:
\begin{lstyaml}
Sampling:
  Method: "Dynesty" 
  Variables:
  - name: x
    description: "Variable X"
    distribution: ... 
       
  - name: y
    description: "Variable Y"
    distribution: ... 

  Bounds:
    nlive:  1600
    rseed:  21
    run_nested: 
      print_progress: True
      maxcall:  500000  
      
  LogLikelihood: 
    - {name: "LogL_Z",  expression: "(2 + z)**5"}
  selection: "(2.0 * X < Y)"
\end{lstyaml}
The grammar of setting up \ykey{Variables}, \ykey{LogLikelihood}, and \ykey{selection} is the same as other variables.
The \code{dynesty} parameters set up in \ykey{Bounds} are introduced as follows\footnote{Documentation for \code{dynesty} can be found here: \href{https://dynesty.readthedocs.io/en/stable/index.html}{https://dynesty.readthedocs.io}.    }:
\begin{itemize}
    \item \ykey{nlive}: Number of live points. Larger numbers indicate a more finely sampled posterior and more accurate evidence. However, this may require more iterations for convergence. 
    \item \ykey{rseed}: Seed for a random number generator. 
    \item \ykey{run\_nested}: Parameters for the dynamic nested sampling loops\footnote{A detailed description of the parameters of the function \yvalue{run\_nested} can be found at: \href{https://dynesty.readthedocs.io/en/stable/api.html\#dynesty.dynamicsampler.DynamicSampler.run_nested}{dynesty.dynamicsampler.DynamicSampler.run\_nested}.}.
    In this category, \ykey{print\_progress} is a flag used to turn on the printing out of a simple summary of each iteration.
    And \ykey{maxcall} is used to set the maximum number of times the likelihood is evaluated.
    If one of the two parameters is not declared in the \code{YAML} interface,
    the default settings will be used. 
\end{itemize}
\ykey{run\_nested} can be used to specify a range of hard stopping criteria\footnote{A pedagogical guide can be found at: \href{https://dynesty.readthedocs.io/en/stable/dynamic.html\#dynamic-nested-sampling}{https://dynesty.readthedocs.io/en/stable/dynamic.html\#dynamic-nested-sampling}. },
which are listed as follows: 
\begin{itemize}
    \item The user can choose \ykey{n\_effective} to set the desired effective number of samples in the posterior. The \code{dynesty} sampler will continue to loop until the \ykey{n\_effective} number of samples is reached. 
    \item The user can specify a maximum number of iterations by choosing \ykey{maxiter} and/or likelihood calls by specifying \ykey{maxcall} in the main loop.
\end{itemize}
In our tests, the \code{dynesty} sampler generated approximately $500,000$ samples based on the specified sampling configurations,
with the results illustrated in Fig.~\ref{fig:dynt} through a scatter plot color-coded according to the likelihood $\mathcal{L}$.
Fig.~\ref{fig:dynesty} demonstrates the detailed behavior of the \code{dynesty} algorithm as it explores the multimodal \code{Eggbox} likelihood landscape.
In the topmost panel, each point represents the remaining prior volume $-\ln(X)$ of a newly generated sample against the cumulative iteration count. Initially, the algorithm progresses swiftly through large prior volumes but subsequently slows as it identifies and focuses on high-likelihood regions.
Directly below, the blue curve indicates the running estimate of the Bayesian evidence, converging toward a stable value of approximately 0.058 as more points are sampled, with the pale green band showing the associated 1$\sigma$ uncertainty range.
The third panel presents the normalized probability density of importance weights, highlighting a prominent peak around $-\ln(X) \approx 3$, where most evidence accumulates, accompanied by a long tail representing the diminishing contributions from samples with smaller prior volumes. The fourth panel illustrates the likelihood evolution,
transitioning from near-zero values in low-likelihood regions to unity as the sampler converges towards regions of higher posterior density.
Finally, the bottom panel tracks the reduction in the number of live points, starting from approximately $1600$ at large prior volumes (small $-\ln(X)$) and gradually decreasing as likelihood contours become tighter,
ultimately reaching zero when the sampling process fully traverses the target distribution (note that the number of live points being maintained stays fixed, but the number that get rejected and replaced at each iteration gradually reduces).
Fig.~\ref{fig:dynesty} effectively demonstrates how \code{dynesty} strikes a balance between broad exploration and precise exploitation, enabling efficient computation of both the Bayesian evidence and posterior samples in complex, multimodal distributions.

\subsection{Machine Learning based Sampling Methods}
Machine learning leverages statistical learning methods to extract patterns from large-scale datasets, employing either supervised or unsupervised approaches.
Machine learning methods have recently garnered significant attention for their application to HEP phenomenological problems.
Our framework employs synchronous calls to HEP packages via a standard worker factory.
Its modular design facilitates machine learning integration at various levels, extending beyond mere improvements in local likelihood evaluation.

The current version of Jarvis-HEP integrates a deep neural network (DNN) sampler, as proposed by Ren et al.~\cite{Ren:2017ymm}.
The DNN sampler employs a DNN to learn physical observables by minimizing a mean squared error (MSE) objective,
which quantifies the discrepancy between the DNN's outputs and theoretical values.
Since the function mapping from parameter space to physical observable space is approximated by DNNs,
we further employ a rejection sampling step~\citep{forsythe1972neumann} to correct any bias introduced by the DNNs,
thereby ensuring the sampled parameters adhere to the target distributions.

The DNN effectively functions as an adaptive proposal mechanism that dynamically learns from the target distribution during the sampling process.
A subsequent rejection sampling step then accepts or rejects these proposals to ensure the sampled parameters accurately adhere to the target distribution.
More specifically, the DNN sampler's workflow proceeds as follows:
A DNN is trained to learn the physical observables corresponding to randomly sampled parameters.
Observables for these points are subsequently computed using HEP packages,
and the results are incorporated into the training set to iteratively refine the model.
This process iterates until a sufficient number of target samples are collected.

To illustrate the DNN sampling method, we employ the \code{Eggbox} potential as an illustrative example.
\begin{lstyaml}
Sampling:
  Method: "DNN" 
  Variables:
    - name: x
      ...
    - name: y
      ... 
  Bounds: 
    Niters: 100
    Hidden_layers: [32, 64, 128, 16]
    Ninit:  200
    Nepoch: 1000
    Learning_rate:  0.0001
    Outputs:  
      - z
      
  LogLikelihood: 
    - {name: "LogL_Z",  expression: "LogGauss(z, 100, 10)"}

  selection: "log(2.0 * x) < sin(y)"
\end{lstyaml}
For this implementation, the \ykey{Method} configuration parameter must be set to \yvalue{DNN}.
Other parameters, specifically \ykey{Variables}, \ykey{LogLikelihood}, and \ykey{selection}, retain their previously defined settings.
The \ykey{Bounds} parameter configures the DNN sampler, which is realized as a customizable feedforward neural network within the PyTorch framework.
This network architecture supports multiple hidden layers featuring ReLU activation functions and optional dropout for regularization.
Its layers are flexibly configurable based on the user-defined architecture.
\begin{itemize}
    \item The input layer is automatically configured based on the specified \ykey{Variables} parameter, which defines the independent variables of the model.
    The output layer is defined by the \ykey{Outputs} parameter,
    comprising a list of physical observables calculated using external packages within the worker factory.
    For the presented Eggbox potential example, the output layer corresponds to \code{z} as defined in Eq.~(\ref{eq:eggbox}),
    while the input consists of the independent variables.
    \item The \ykey{Hidden\_layers} parameter is specified as a list, where each numerical entry denotes the count of neurons within the corresponding hidden layer.
    \item The \ykey{Niters} parameter denotes the total number of training iterations, or parameter updates.
    Each iteration typically constitutes a single update step performed with one batch of data. In this context, the model undergoes 40 such update steps in total.
    \item The \ykey{Ninit} parameter determines the number of samples generated within each iteration.
    \item The \ykey{Nepoch} parameter specifies the number of epochs for each training phase. An epoch signifies a complete pass of the entire training dataset through the neural network. Employing multiple epochs generally facilitates enhanced model learning from the data.
    \item The \ykey{Learning\_rate} parameter governs the magnitude of weight adjustments in the model with respect to the gradient. A small learning rate, such as $10^{-4}$ in this case, typically leads to slower but more stable convergence, making it particularly suitable for sensitive or complex models.
    \item The \ykey{Prop\_new} parameter influences the sampling process within each iteration:
    a large batch of randomly generated samples (equal to \ykey{Ninit}/\ykey{Prop\_new}) is initially pre-sampled, from which the DNN then prioritizes those with higher likelihoods according to a specified ratio. The default value for \ykey{Prop\_new} is $0.1$.
\end{itemize}
\begin{figure}
	\centering\hspace{-0.3cm}
	\includegraphics[width=0.48\textwidth]{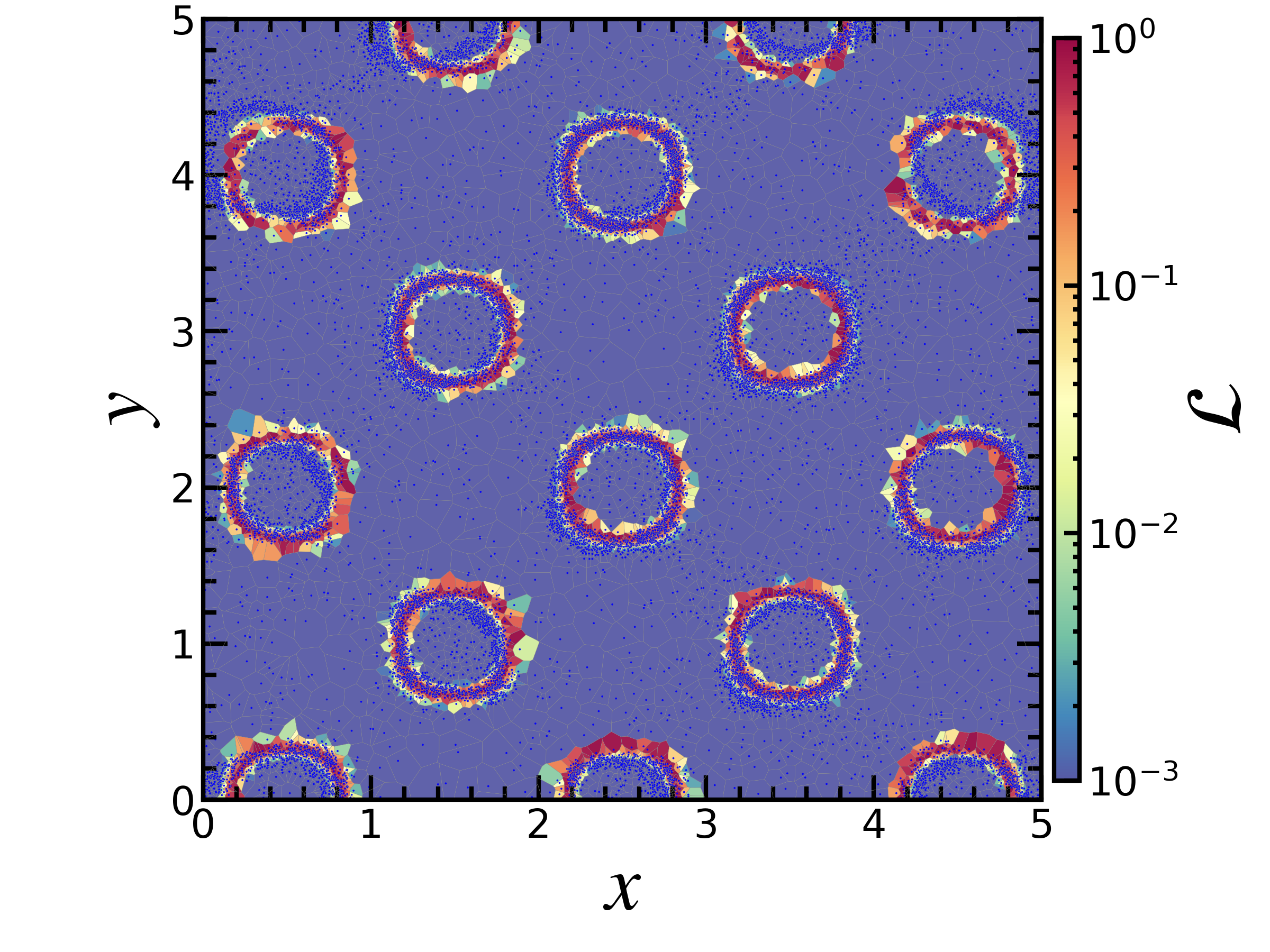}
	\caption{\label{fig:dnn} Similar to Fig.~\ref{fig:simsam}, but using DNN sampling method.}
\end{figure}
The total samples in the above sampling is $20000$ and the result is shown in Fig.~\ref{fig:dnn}. One can find that a large proportion of samples are distributed in places with a large likelihood, and all peaks have been found.
Designed for flexible implementation of diverse sampling algorithms, \code{Jarvis-HEP} aims to evolve with user needs,
focusing future efforts on integrating advanced AI and addressing inherent challenges like checkpointing in asynchronous execution.


\section{Example: deploy scan tasks and understand the results}
\label{sec:example}
\par This section uses the \code{EggBox} project as a concrete example to demonstrate how a scan task is deployed and inspected in \code{Jarvis-HEP}. 
The focus is on practical usage: fetching a ready-to-run project, launching a scan from the command line, following the execution through logs and monitoring tools, understanding the generated output directories, and producing basic visualisations from the recorded results.
\par This section guides users through the typical workflow from task execution to result inspection. 
It provides a practical bridge between the configuration interface described in Sec.~\ref{sec:yaml} and the use of \code{Jarvis-HEP} in real numerical studies.

\subsection{Setup: fetching and inspecting the example project}
\label{sec:example-setup}

The \code{EggBox} example is distributed as a ready-to-run \code{Jarvis-HEP} project. 
It can be obtained directly from the command line by
\begin{lstterm}
Jarvis project fetch EggBox
\end{lstterm}
This command downloads an already configured project from the repository and places it in a local project directory, including the YAML configuration files, the lightweight calculator used in the example, and the auxiliary files required to reproduce the workflow discussed in this section.

\par After fetching the project, the main directory structure is organised as follows:
\dirtree{%
.1 EggBox/.
.2 bin/.
.3 Example\_Dynesty.yaml.
.3 Example\_Random.yaml.
.3 ....
.3 Example\_Grid.yaml.
.2 deps/.
.3 EggBox/.
.3 environment\_default.yaml.
.2 README.md/.
.2 jarvis.project.yaml/. 
}
The \file{bin/} directory contains the YAML entry points used to launch scan tasks. 
The \file{deps/} directory stores the lightweight calculator and auxiliary files used by the example.
The \file{README.md} file provides a short description of the example project and records the basic commands needed to reproduce the workflow.
The file \file{jarvis.project.yaml} records project-level metadata and allows \code{Jarvis-HEP} to identify the directory as a project workspace.

\subsection{Running and monitoring a scan task}
\label{sec:example-run-monitor}

After the example project has been fetched and inspected, a scan task can be launched by passing one of the YAML entry files to the \code{Jarvis} command. 
For the nested-sampling example used in this section, the task is started from the project root by
\begin{lstterm}
Jarvis ./bin/Example_Dynesty.yaml
\end{lstterm}
During startup, \code{Jarvis-HEP} parses the YAML file, resolves project-relative paths, prepares the calculator workspace, constructs the task workflow, and starts the sampler together with the worker processes.
\par Once the task starts, \code{Jarvis-HEP} creates the runtime directories for this scan under the project workspace. 
For the \file{Example\_Dynesty.yaml} configuration, the generated structure is schematically shown below:
\dirtree{%
.1 EggBox/.
.2 images/EggBox\_Dynesty/.
.2 outputs/EggBox\_Dynesty/.
.4 DATABASE/.
.4 SAMPLE/.
.2 logs/EggBox\_Dynesty/.
.2 checkpoints/EggBox\_Dynesty/.
.2 calculators/.
}
Here, the directories \file{images/}, \file{outputs/}, \file{logs/}, and \file{checkpoints/} are organised by scan name, so that files generated by different scan tasks remain separated within the same project. For this example, the scan name is \file{EggBox\_Dynesty}. 
The \file{calculators/} directory contains multiple installed copies of the calculator modules used by the workers to execute tasks in parallel.

\subsubsection{Logging system}
\label{sec:example-logging}
\code{Jarvis-HEP} dispatches log messages to three distinct outputs:
\begin{itemize}
  \item Terminal (console): Displays essential runtime information and provides a real-time view of the current status of the scan task.
  \item Log files: Stored under the scan-specific directory, \file{logs/EggBox\_Dynesty/} in this example, with separate log files used to record messages from the sampler, the worker factory, and the main \code{Jarvis} scan process. These logs allow users to inspect the sampling procedure, task scheduling, calculator execution, and overall scan status independently.
  \item Sample log files: Point-level logs are stored under the scan-specific logging directory, for example in \file{outputs/EggBox\_Dynesty/SAMPLE/Sample-UUID/Sample\_running.log}. Each file records the execution history of an individual sampled point, including the calculator modules called by the worker, the input and output handling steps, and any errors raised during execution. These logs are useful for debugging failed points and for inspecting the detailed runtime history of specific samples.
\end{itemize}
This logging design is particularly useful for multi-threaded scans, where many workers call external programs and system commands simultaneously. 
\code{Jarvis-HEP} automatically organises these outputs into structured log files, allowing users to analyse the sampling behaviour, inspect individual task histories, and diagnose failures after the scan has finished.

During execution, \code{Jarvis-HEP} writes log files under a scan-specific logging directory. 
For the \code{EggBox\_Dynesty} example, the logs are stored in
\begin{lstterm}
logs/EggBox_Dynesty/
\end{lstterm}
The log records the main stages of the workflow, including YAML parsing, environment checks, calculator preparation, task submission, worker execution, data recording, error messages, and final termination.

\par To help users monitor running tasks and locate issues after execution, \code{Jarvis-HEP} writes structured log messages both to the terminal and to log files. 
Each message follows a common format:
\begin{lstterm}
Jarvis-HEP.Module 
    -> TIME STAMP - [LEVEL] >>> 
  MESSAGE BODY 
\end{lstterm}

\par This format is designed to make runtime information human-readable while still preserving enough context for debugging.
The prefix identifies the component that generated the message, such as \code{Jarvis-HEP.ConfigParser}, and \code{Jarvis-HEP.Factory}. 
The timestamp records when the message was produced, using the format \code{MM-DD HH:MM:SS.sss}. 
The log level, such as \code{[INFO]}, \code{[WARNING]}, or \code{[ERROR]}, indicates the severity of the message. 

\subsubsection{Checking modules before a full scan}
\label{sec:example-check-modules}
Before launching a full scan, users can test whether the YAML configuration and calculator workflow can be executed correctly using the module-checking mode:
\begin{lstterm}
Jarvis ./bin/Example_Dynesty.yaml --check-modules
\end{lstterm}
In this mode, \code{Jarvis-HEP} starts a minimal scan with 10 sampled points. 
The purpose is to verify that the YAML file can be parsed, the calculator modules can be prepared, the required input and output files can be handled correctly, and the likelihood evaluation can be completed for representative samples.

\par The \code{--check-modules} option is intended for debugging and validation rather than for producing physics results. When this mode is enabled, the terminal log level is automatically set to debug, so that users can see more detailed runtime information directly from the console. 
This makes it easier to identify problems such as missing files, incorrect paths, failed external commands, or errors in symbolic expressions before starting a long scan.

\subsubsection{Monitoring a running scan}
\label{sec:example-monitor}
While a scan is running, users can monitor its computing resources in separate console using
\begin{lstterm}
Jarvis ./bin/Example_Dynesty.yaml --monitor
\end{lstterm}

\begin{figure}[th]
	\includegraphics[width=0.98\linewidth]{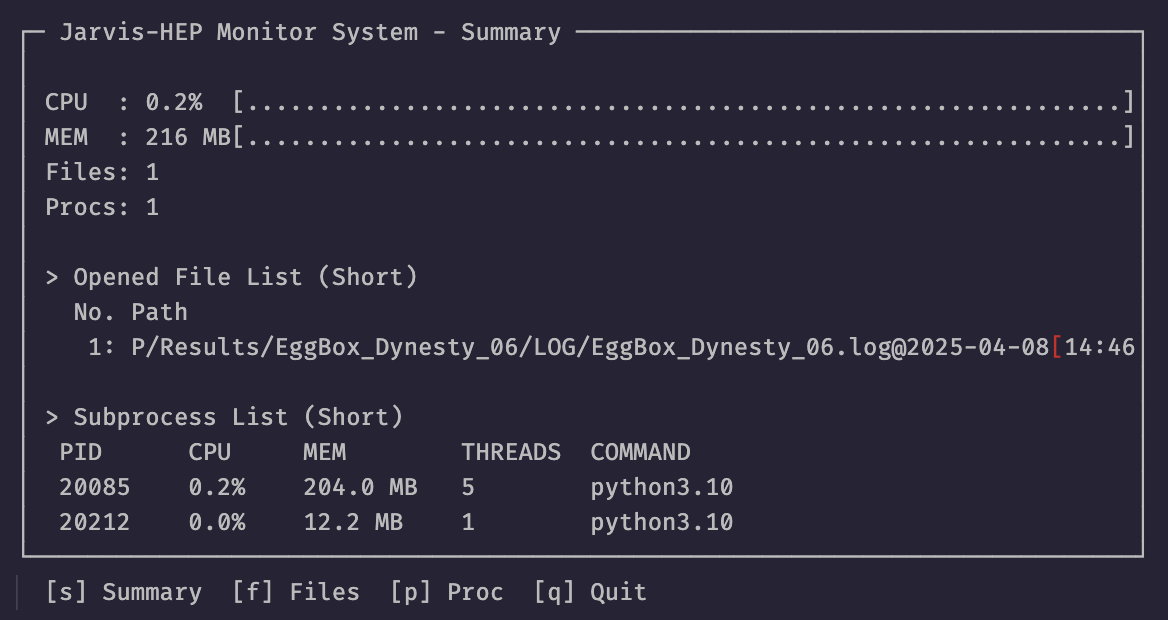}
	\caption{\label{fig:monitor}Live computing resources monitor in \code{Jarvis-HEP}.}
\end{figure}
Then \code{Jarvis-HEP} starts a text user interface application, as shown in Fig.~\ref{fig:monitor}. This monitor is particularly useful for tracking the runtime behavior of parallel scans or computationally expensive factory evaluations. The interface includes:

\begin{itemize}
  \item CPU and memory load: shown both numerically and as a visual bar, allowing users to quickly assess whether the system is under stress.
  \item Opened file list: display currently opened files, including the number index and the full path. If the called program has abnormal exit behaviour, you can see an unclosed file in some cases. In addition, for the smooth operation of the system, the number of files that a user can open at the same time is limited, so users can monitor here.
  \item Subprocess list: lists all active subprocesses spawned by the Jarvis engine, including their PID, CPU and memory usage, number of threads, and the command used.
\end{itemize}
The bottom command bar offers interactive navigation shortcuts, such as \code{[s]} for the summary view, \code{[f]} for the opened file list, and \code{[p]} for the detailed process view. This lightweight but informative monitor makes \code{Jarvis-HEP} especially suitable for long scans on remote servers or headless environments, providing a convenient way to detect bottlenecks, debug misbehaving modules, or monitor progress at a glance. This monitor does not affect the main scanning process and can be turned on and off at any time.

\subsubsection{Converting recorded data during execution}
\label{sec:example-convert}
During the execution of a scan task, \code{Jarvis-HEP} records scan data in \code{HDF5} format, enabling efficient and reliable writing by the data recorder.
To inspect the accumulated results while the scan is still running, users can convert the recorded data into \code{CSV} format from a separate console:
\begin{lstterm}
Jarvis ./bin/Example_Dynesty.yaml --convert
\end{lstterm}
This mode safely reads the current \code{HDF5} dataset and produces a \code{CSV} file for intermediate data inspection, without interrupting the running scan or modifying the active recording file.

\subsubsection{Checkpointing and resuming an interrupted scan}
\label{sec:example-checkpoint}

During execution, \code{Jarvis-HEP} periodically writes checkpoint files for the running scan. 
For the \code{EggBox\_Dynesty} example, these files are stored under the scan-specific checkpoint directory:
\begin{lstterm}
checkpoints/EggBox_Dynesty/
\end{lstterm}
The checkpoint files record the runtime state needed to continue an interrupted task, including the sampler state and the progress of the scan workflow.

\par If a scan is interrupted before completion, users can restart the same command from the project root:
\begin{lstterm}
Jarvis ./bin/Example_Dynesty.yaml
\end{lstterm}
When checkpoint files are found, \code{Jarvis-HEP} can resume the scan from the saved state instead of starting from the beginning. 
This is useful for long-running scans, where interruptions may occur because of terminal disconnection, system shutdown and time limits.

\par The checkpoint directory is organised by scan name, so different scan tasks in the same project maintain independent checkpoint states. 
Users should keep the scan name and output paths unchanged when resuming a task, since these settings determine where \code{Jarvis-HEP} searches for the saved checkpoint files.

\subsubsection{Run summary}
\label{sec:example-run-summary}
When a scan finishes, \code{Jarvis-HEP} prints a run summary to the terminal and records it in the log file. 

The summary reports the run identifier, wall time, worker usage, submitted and completed points, failed tasks, success rate, throughput, and basic resource statistics. 
A shortened example is shown below:
\begin{lstterm}
[Run Overview]
+----------------------+------------------+
| Metric               | Value            |
+----------------------+------------------+
| Run ID               | EggBox_Dyne-...  |
| Sampler              | Dynesty          |
| Wall Time (s)        | 1425.919         |
| Workers Configured   | 16               |
| Peak Active Workers  | 16               |
| Points Submitted     | 765536           |
| Points Completed     | 765536           |
| Points Failed        | 0                |
| Success Rate         | 100.00
| Throughput (pts/min) | 32212.311        |
+----------------------+------------------+
\end{lstterm}

\par This summary allows users to quickly verify whether the scan completed successfully and whether the configured workers were effectively used. 
The number of failed points and the success rate provide a direct check of runtime stability, while the throughput and worker-usage information help users evaluate the practical efficiency of the run. 
The full summary also includes an execution breakdown with point-level timing, CPU usage, memory usage, and open-file statistics, which can help diagnose bottlenecks in larger scans.

\par In repeated tests on a typical personal computer, \code{Jarvis-HEP} was able to sustain a throughput of approximately $3\times 10^4$--$6\times 10^4$ completed lightweight tasks per minute. 
This indicates that the asynchronous, non-blocking scheduling layer introduces only a small overhead compared with the execution of the calculator modules. 
For most compute-intensive scan applications, where the runtime is dominated by external physics calculations, this scheduling capacity is sufficient for practical large-scale parameter scans.

\subsection{Understanding the output structure}
\label{sec:example-output-structure}
For the present example, the structure of the output directory is
\dirtree{%
.1 outputs/EggBox\_Dynesty/.
.2 run\_summary.txt. 
.2 DATABASE/.
.3 archive\_manifest.jsonl.
.3 running.json.
.3 samples.0.csv.
.3 samples.0.hdf5.
.3 samples.schema.json. 
.2 SAMPLE/.
.3 000001.tar.gz. 
.3 000002.tar.gz. 
.3 ....
.3 003828.tar.gz. 
}
\par The \file{DATABASE/} directory is the global database directory of a scan: 
\begin{itemize}
	\item The file \file{samples.0.hdf5} is the first HDF5 data shard written by the global recorder.
Although the configured database root is \file{DATABASE/samples.hdf5}, the data are written to numbered shards such as \file{samples.0.hdf5} to avoid creating an overly large single data file.
Each shard stores JSON-like sample records containing sampled parameters, calculator outputs, likelihood values, and other recorded quantities. 
	\item The corresponding file \file{samples.0.csv} is CSV exports of these HDF5 shards, intended for quick inspection and plotting with common data-analysis tools.
 	\item The file \file{samples.schema.json} records the schema of the sampled data, including field types, array or list structures, dictionary-like entries, and the flattening rules used when exporting complex records to CSV. 
 	\item The file \file{running.json} stores lightweight runtime metadata for the database writer, such as the active HDF5 shard, shard index, converted CSV files, historical shards, and error information. 
	\item The file \file{archive\_manifest.jsonl} records the archive manifest of sample buckets, for example \file{SAMPLE/000001.tar.gz}. Each line describes an archived \file{SAMPLE} bucket, including its source path, archive path, and timestamp. 
\end{itemize}
The \file{SAMPLE/} directory stores point-level records associated with sampled parameter points. 
To avoid creating a large number of small files, \code{Jarvis-HEP} groups sampled points into archive buckets; by default, each bucket contains 200 samples. 
A typical bucket has the following structure:
\dirtree{%
.1 outputs/EggBox\_Dynesty/SAMPLE/000001/. 
.2 Sample-A-UUID/. 
.3 Sample\_running.log.
.3 output.json@EggBox.
.2 Sample-B-UUID/. 
.2 ....
.2 Sample-Z-UUID/. 
}
Each extracted bucket contains several sample-specific subdirectories, identified by their sample UUIDs. 
For each sampled point, files such as \file{Sample\_running.log} record the point-level execution history, while files such as \file{output.json@EggBox} store the calculator output produced by the corresponding module. 
\par The separation between \file{DATABASE/} and \file{SAMPLE/} reflects two complementary views of the scan result. 
The former provides a compact tabular dataset suitable for statistical analysis and visualisation, while the latter keeps detailed per-sample information for inspection and reproducibility. 

\subsection{Data visualisation}
\label{sec:example-data-visualisation}
After the scan data have been recorded, the next step is to visualise the sampled points and the corresponding likelihood values. 
\par The visualisation mode uses the CSV version of the recorded dataset as input and generates a plotting YAML configuration for the companion tool \code{Jarvis-PLOT}\footnote{\code{JarvisPLOT} is now available on PyPI. See \url{https://pypi.org/project/jarvisplot/}. }. 
The plotting configuration is constructed from the variables and likelihood quantities defined in the scan configuration.

\par A basic visualisation can be generated from the project directory by running
\begin{lstterm}
Jarvis ./bin/Example_Dynesty.yaml --plot
\end{lstterm}
This command generates the \code{Jarvis-PLOT} configuration file used to produce the figures stored in the scan-specific image directory, located at \file{images/EggBox\_Dynesty/EggBox\_dynesty.yaml}. 
Users can then modify the generated drawing profile if needed and produce the figures using the \code{jplot} command:
\begin{lstterm}
jplot images/EggBox_Dynesty/EggBox_dynesty.yaml
\end{lstterm}
The command \code{jplot} is provided by the companion package \code{Jarvis-PLOT} and becomes available after the package is installed via \code{pip}. 
This command generates the visualisation results shown in Figs.~\ref{fig:dynt} and Fig.~\ref{fig:dynesty}. The visualisation step provides a direct way to inspect the sampled parameter space.

\subsection{Packing the project for deployment and archiving}
\label{sec:example-project-pack}

After a scan workflow has been configured and tested, the project can be packaged into a portable bundle using
\begin{lstterm}
Jarvis project pack [path] [--share | --repro | --full]
\end{lstterm}
Here, \code{[path]} specifies the project directory to be packed. If it is omitted, the current project directory is used. The command supports three packing modes:
\begin{itemize}
  \item \code{--share}: creates a lightweight bundle suitable for sharing the project structure, configuration files, and essential resources. This is the default mode when no option is specified.
  \item \code{--repro}: creates a reproducible bundle intended to preserve the information needed to rerun or verify the workflow.
  \item \code{--full}: creates a full bundle containing the most complete project snapshot, including generated files and results.
\end{itemize}

\par For example, a lightweight shareable bundle can be created by
\begin{lstterm}
Jarvis project pack
\end{lstterm}
or explicitly by
\begin{lstterm}
Jarvis project pack . --share
\end{lstterm}

\par The packed project can be transferred to another machine, shared with collaborators, or stored as a reproducible snapshot of the scan workflow. 
Together with the project-fetching command used at the beginning of this section, this provides a simple project-level cycle: a project can be fetched, executed, inspected, visualised, and packed again for deployment or long-term archiving.

\subsection{Practical notes}
\label{sec:example-faq}

This section summarises several practical questions that may arise when running the example project.
\begin{enumerate}
\item \textit{Can installed dependencies and calculator modules be reused?}
\par Yes. \code{Jarvis-HEP} records configuration information for installed dependencies and calculator modules, and reuses existing installations when the current configuration is compatible. 
After the workflow has been validated with \code{--check-modules}, users can usually launch production scans directly. 
Changing only the sampling method does not normally require reinstalling external packages or calculator modules.

\item \textit{Where should users check if a scan fails?}
\par Users should first inspect the scan-specific log directory, such as \file{logs/EggBox\_Dynesty/}. 
The YAML check mainly validates syntax, so users should also examine the generated workflow graph to verify the intended computational dependencies. 
For detailed diagnosis, point-level logs in the sample records can be used to trace calculator calls, system commands, input/output handling, and likelihood evaluations.

\item \textit{Can the generated plots be modified?}
\par Yes. Users can edit the generated plotting YAML file and rerun it with \code{jplot}. 
Custom plotting syntax is documented in the independent companion project \code{Jarvis-PLOT}.

\item \textit{Can \code{Jarvis-HEP} read values directly from screen output?}
\par No. Screen output is recorded in the log files for debugging, but it is not used as a structured source of numerical observables. 
If an external program prints the required quantities to the terminal, users should redirect the output to a file or wrap the program with a post-processing script that converts the relevant values into a format recognised by \code{Jarvis-HEP}, such as \code{JSON}. 
For commonly used packages, users are encouraged to report missing output-handling interfaces, since \code{Jarvis-HEP} is developed in response to practical user requirements.

\item \textit{Must all external dependencies be managed by \code{Jarvis project}?}
\par No. External programs do not need to be installed through the project-management interface. 
They can be located in \file{WorkShop/}, \file{calculators/}, system paths, or any user-defined directory. 
\code{Jarvis-HEP} only requires the paths, optional setup commands, execution commands, and input/output definitions needed to call them. 
In principle, any command that runs correctly in the user's terminal can be connected to a \code{Jarvis-HEP} workflow.

\item \textit{How should environment variables be set for external commands?}
\par \code{Jarvis-HEP} does not currently provide a dedicated \code{env:} field for commands. 
The \code{EnvReqs} block checks environment requirements, but does not set variables such as \code{CC}, \code{CXX}, \code{PATH}, or \code{LD\_LIBRARY\_PATH}. 
Since each command is executed independently, required environment settings should be included in the same shell command, for example:
\begin{lstterm}
CC=gcc CXX=g++ make -j4
source /opt/root/bin/thisroot.sh && ./run input.dat output.dat
\end{lstterm}
For complex setups, users are encouraged to use a wrapper script that prepares the environment and then calls the external program.

\end{enumerate}

\section{Summary and future directions}
\label{sec:sum}

In this work, we have presented \code{Jarvis-HEP}, a Python framework for composing and executing parameter-scan workflows in high-energy physics. 
Rather than focusing on a single sampling algorithm, \code{Jarvis-HEP} provides a project-based workflow layer that connects YAML-defined configurations, external calculator packages, likelihood evaluation, asynchronous task execution, structured data recording, logging, monitoring, visualisation, and project packaging.

\par The framework is designed to help researchers combine multiple computational tools in a transparent and reproducible way. 
A scan workflow can be described in a single YAML file, including variables, likelihoods, environment checks, dependencies, calculator modules, and input/output handling. 
The worker-factory design then converts this configuration into an executable computation graph, while the asynchronous execution model allows multiple tasks to be processed concurrently. 
Together with structured logging and HDF5-based data recording, this provides both machine-readable scan results and a detailed record of runtime behaviour.

\par The examples in this paper demonstrate the basic sampling interfaces and the end-to-end use of the framework. 
In particular, the \code{EggBox} project illustrates a complete workflow, including project fetching, module checking, scan execution, monitoring, data conversion, output inspection, visualisation through \code{Jarvis-PLOT}, and project packing for deployment or archiving.

\par \code{Jarvis-HEP} should be viewed as a lightweight and extensible workflow framework rather than a replacement for large global-fitting ecosystems. 
Its main strength lies in rapid deployment, transparent project organisation, and flexible integration of external command-line tools. 
This makes it useful for exploratory studies, method development, reproducible numerical workflows, and small-to-medium scale parameter scans.

\par Future development will focus on extending supported input/output formats, improving interfaces to commonly used high-energy physics packages, adding further sampler backends, and improving interoperability with companion Python packages in the \code{Jarvis} ecosystem.
In addition, we will also explore AI-agent-assisted workflow construction, aiming to support the integration, migration, and validation of complex computational pipelines in \code{Jarvis-HEP}. Users are encouraged to combine AI assistance with the official documentation and manual checks when preparing YAML configurations and designing scan tasks.
We also plan to expand the collection of public example projects and documentation, with user feedback continuing to guide the evolution of the framework.

\section*{Acknowledgement}
\addcontentsline{toc}{section}{Acknowledgement}
We thank Martin J. White for his valuable contributions to the code development, the design of the scanning algorithms and statistical framework, and the preparation of this manuscript.
PZ thanks Jonathan Woithe for valuable discussions on parallel programming. 
PJ is supported by the ARC Centre of Excellence for Dark Matter Particle Physics CE200100008. 
PJ and PZ are supported by the ARC Discovery Project DP220100007. 
PZ is supported by the Centre for the Subatomic Structure of Matter (CSSM).
JMY is supported by the National Natural Science Foundation of China under Grant No. 12335005,  by the Peng-Huan-Wu Theoretical Physics Innovation Center funded by the National Natural Science Foundation of China (Grant No.12447101) and by the Research Fund for PI from Henan Normal University under Grant No. 5101029470335. 

\newpage
\appendix
\section{Installation and Dependencies}
\label{app:ins}
\code{Jarvis-HEP} is implemented in pure Python and has been tested on Unix-like operating systems (Linux, macOS). 
\subsection{Installation via pip (PyPI)}
The fundamental requirement is \code{Python-3.10} or a more recent version. One can use virtual environments (e.g., \code{Conda} or \code{pyenv}) to maximize reproducibility and portability.  
You can install Jarvis-HEP using pip, the package management tool for Python:
\begin{lstterm}
pip install Jarvis-HEP
\end{lstterm}
or equivalently via the Python module interface:
\begin{lstterm}
python -m pip install Jarvis-HEP
\end{lstterm}
After installation, \code{pip} will provide a command-line interface \code{Jarvis}. You can verify it by running:
\begin{lstterm}
Jarvis -h
\end{lstterm}

\subsection{Code Repository and Documentation}
\code{Jarvis-HEP} is openly hosted on GitHub at
\begin{center}
\href{https://github.com/Pengxuan-Zhu-Phys/Jarvis-HEP}{https://github.com/Pengxuan-Zhu-Phys/Jarvis-HEP}.
\end{center}
The repository contains the full source code, issue tracker, and development history, and serves as the primary entry point for reference and contributions.

The official documentation of \code{Jarvis-HEP}, covering installation, workflows, and API usage in detail, is available at
\begin{center}
\href{https://pengxuan-zhu-phys.github.io/Jarvis-Docs/}{https://pengxuan-zhu-phys.github.io/Jarvis-Docs/}.
\end{center}
It is regularly maintained to keep pace with ongoing development and provides a unified reference for the entire \code{Jarvis} ecosystem, including \code{Jarvis-PLOT} for visualization and \code{Jarvis-Operas}, an operator layer for the unified registration and invocation of Python functions with support for synchronous/asynchronous execution and extensibility.

\subsection{Mandatory Dependencies}
During installation, \code{pip} will automatically resolve and install all required dependencies with compatible versions to ensure that \code{Jarvis-HEP} functions correctly. Users are therefore not required to manually install individual packages. In case of environment inconsistencies or missing dependencies, reinstalling or upgrading \code{Jarvis-HEP} via \code{pip} is usually sufficient to restore a consistent setup.

For completeness, the main dependencies used by \code{Jarvis-HEP} are listed below:
\begin{itemize}
	\item \code{NumPy}~\cite{vanderWalt:2011bqk, harris2020array}: for numerical array operations.
	\item \code{Scipy}~\cite{2020SciPy-NMeth}: for scientific computing routines (integration, optimization, special functions, etc.).
	\item \code{Matplotlib}~\cite{Hunter:2007}: for visualization and graphical output.
	\item \code{dynesty}~\cite{Higson_2018, Feroz:2008xx}: for dynamic nested sampling, used for Bayesian inference and computing high dimensional integrals efficiently. 
	\item \code{pandas}~\cite{reback2020pandas}: for data handling and analysis.
	\item \code{PySLHA}~\cite{Buckley:2013jua}: for reading and writing supersymmetric model data in the standard SUSY Les Houches Accord (SLHA) format~\cite{Skands:2003cj, Allanach:2008qq}.
	\item \code{xslha}~\cite{Staub:2018rih}: for supporting non-standard SLHA files. See also in Sec.~\ref{sec:yaml}. 
	\item \code{xmltodict}: for working with I/O files with extensible markup language (XML) format. 
	\item \code{aiofiles}: for asynchronous file operations with \code{asyncio} support.
	\item \code{dill}: for serialising and de-serialising Python objects. 
	\item \code{h5py}: for storing huge amounts of numerical data. 
	\item \code{PyYAML}: for parsing the human friendly input YAML file. 
	\item \code{jsonschema}: for validating the input YAML file against a defined JSON data schema, ensuring the YAML adheres to specific structural and content rules.
	\item \code{loguru}: for logging output.
	\item \code{networkx}~\cite{SciPyProceedings_11}: for parsing the logical relationship of calling packages. 
	\item \code{PrettyTable}: for formatting table screen output. 
	\item \code{psutil}: for retrieving information on running processes and system utilisation (CPU, memory and number of opened files). 
	\item \code{shapely}~\cite{shapely}: for handling geometric operations and spatial data processing and visualisation. 
	\item \code{sympy}~\cite{sympy:2017}: for symbolic mathematics, enabling the analytical manipulation, simplification, and evaluation of mathematical expressions. 
	\item \code{emoji}: for enhancing user interactions and the logging system. 
	\item \code{PyTorch}~\cite{pytorch:2019}: a popular machine learning library. 
\end{itemize}

\renewcommand{\thetable}{A\arabic{table}} 
\section{Symbolic Expressions in Jarvis-HEP}
\label{app:expr}

\begin{table}[h!]
\caption{List of supported mathematical symbols in \code{Jarvis-HEP}. Each constant is represented with its symbolic notation and/or numerical value.}
\renewcommand{\arraystretch}{1.4} 
\label{tab:constants}
\centering
\adjustbox{max width=0.49\textwidth}{%
\begin{tabular}{P{2.9cm}|C{1.8cm}|P{3.6cm}}
\Xhline{1.5pt}
\textbf{Name} & \textbf{Symbol} & \textbf{Description} \\ \Xhline{1.5pt}
\code{Pi} & $\pi$ & the mathematical constant, $\pi \approx 3.141592653589793$ \\  
\code{E} & $e$ & the mathematical constant, $e \approx 2.718281828459045$ \\  
\code{Inf} & $\infty$ & Infinity defined in \code{numpy} \\ \Xhline{1.5pt}
\end{tabular}
}
\end{table}

\begin{table}[h!]
\renewcommand{\arraystretch}{1.3} 
\caption{List of supported mathematical functions in \code{Jarvis-HEP}. Each function is represented using its symbolic form and accompanied by a description of its purpose or mathematical significance.}
\label{tab:functions}
\adjustbox{max width=0.49\textwidth}{%
\begin{tabular}{P{2.9cm}|C{1.8cm}|P{3.6cm}}
\Xhline{1.5pt}
\bf{Name} 						& \bf{Expression} 			& \bf{Description} \\ 
\Xhline{1.5pt}
\code{log(\vars{x},\vars{a})} 	& $\log_{\vars{a}}(\vars{x})$ & General logarithm of $\vars{x}$ in base of \vars{$a$}. \\  
\code{exp(\vars{x})} 	& $e^{\vars{x}}$  & Exponential function. \\  
\code{ln(\vars{x})} 	& $\ln{\vars{x}}$ & Natural logarithm of \vars{$x$}. \\  
\code{sqrt(\vars{x})} 	& $\sqrt{\vars{x}}$ & Square root of $\vars{x}$. \\  
\code{root(\vars{x},\vars{n})} 	& $\sqrt[\vars{n}]{\vars{x}}$ & \vars{$n$}th root of \vars{$x$}.\\
\code{sin(\vars{x})} 	& $\sin(\vars{x})$ & Sine of $\vars{x}$. \\  
\code{cos(\vars{x})} 	& $\cos(\vars{x})$ & Cosine of $\vars{x}$. \\  
\code{tan(\vars{x})} 	& $\tan(\vars{x})$ & Tangent of $\vars{x}$. \\  
\code{sec(\vars{x})} 	& $\sec(\vars{x})$ & Secant of $\vars{x}$. \\  
\code{csc(\vars{x})} 	& $\csc(\vars{x})$ & Cosecant of $\vars{x}$. \\  
\code{cot(\vars{x})} 	& $\cot(\vars{x})$ & Cotangent of $\vars{x}$. \\  
\code{asin(\vars{x})} 	& $\sin^{-1}(\vars{x})$ & Inverse sine of $\vars{x}$. \\  
\code{acos(\vars{x})} 	& $\cos^{-1}(\vars{x})$ & Inverse cosine of $\vars{x}$. \\  
\code{atan(\vars{x})} & $\tan^{-1}(\vars{x})$ & Inverse tangent of $\vars{x}$. \\  
\code{asec(\vars{x})} & $\sec^{-1}(\vars{x})$ & Inverse secant of $\vars{x}$. \\  
\code{acsc(\vars{x})} & $\csc^{-1}(\vars{x})$ & Inverse cosecant of $\vars{x}$. \\  
\code{acot(\vars{x})} & $\cot^{-1}(\vars{x})$ & Inverse cotangent of $\vars{x}$. \\  
\code{atan2(\vars{y},\vars{x})} & $\tan^{-1}\left(\vars{y}/\vars{x}\right)$ & Both \vars{$x$} and \vars{$y$} are real numbers. The range is $\left(-\pi, \pi \right]$. \\  
\code{sinh(\vars{x})} & $\sinh(\vars{x})$ & Hyperbolic sine of $\vars{x}$. \\  
\code{cosh(\vars{x})} & $\cosh(\vars{x})$ & Hyperbolic cosine of $\vars{x}$. \\  
\code{tanh(\vars{x})} & $\tanh(\vars{x})$ & Hyperbolic tangent of $\vars{x}$. \\ 
\code{sech(\vars{x})} & $\operatorname{sech}(\vars{x})$ & Hyperbolic secant of $\vars{x}$. \\ 
\code{csch(\vars{x})} & $\operatorname{csch}(\vars{x})$ & Hyperbolic cosecant of $\vars{x}$. \\  
\code{coth(\vars{x})} & $\operatorname{coth}(\vars{x})$ & Hyperbolic cotangent of $\vars{x}$. \\  
\code{asinh(\vars{x})} & $\sinh^{-1}(\vars{x})$ & Inverse hyperbolic sine of $\vars{x}$. \\  
\code{acosh(\vars{x})} & $\cosh^{-1}(\vars{x})$ & Inverse hyperbolic cosine of $\vars{x}$. \\  
\code{atanh(\vars{x})} & $\tanh^{-1}(\vars{x})$ & Inverse hyperbolic tangent of $\vars{x}$. \\  
\code{acoth(\vars{x})} & $\coth^{-1}(\vars{x})$ & Inverse hyperbolic cotangent of $\vars{x}$. \\  
\code{asech(\vars{x})} & $\operatorname{sech}^{-1}(\vars{x})$ & Inverse hyperbolic secant of $\vars{x}$. \\  
\code{acsch(\vars{x})} & $\operatorname{csch}^{-1}(\vars{x})$ & Inverse hyperbolic cosecant of $\vars{x}$. \\   
\code{Gauss(\vars{x},\vars{$\mu$},\vars{$\sigma$})} & 
 $\exp \left({-\frac{(\vars{x}-\vars{\mu})^2}{2\vars{\sigma}^2}}\right)$ & 
A short Gaussian distribution function. \\  
\code{Normal(\vars{x},\vars{$\mu$},\vars{$\sigma$})} & $N\left(\vars{\mu}, \vars{\sigma}^2 \right)$ & The standard normal distribution function or Gaussian distribution. \\  
\code{LogGauss(\vars{x},\vars{$\mu$},\vars{$\sigma$})} & $-\dfrac{(\vars{x}-\vars{\mu})^2}{2 \vars{\sigma}^2}$ & Logarithmic Gaussian function. \\  
\code{Heaviside(\vars{x})} & $H(\vars{x})$ & Heaviside step function. \\  
\code{Min(\vars{x},\vars{y})} & $\min(\vars{x}, \vars{y})$ & Minimum of $\vars{x}$ and $\vars{y}$. \\  
\code{Max(\vars{x},\vars{y})} & $\max(\vars{x}, \vars{y})$ & Maximum of $\vars{x}$ and $\vars{y}$. \\  
\code{Abs(\vars{x})} & $|\vars{x}|$ & Absolute value of $\vars{x}$. \\ \Xhline{1.5pt}
\end{tabular}
}
\end{table}
\code{Jarvis-HEP} supports the use of symbolic string expressions to represent mathematical expressions that comply with Python syntax, enabling users to define and evaluate complex sample filter conditions or calculations dynamically. This feature provides flexibility for integrating user-defined expressions with Jarvis-HEP's computational modules.

Some common mathematical symbols are summarised in Table~\ref{tab:constants}, and a variety of pre-defined functions are summarised in Table~\ref{tab:functions}. As in Sec.~\ref{sec:utils}, user-defined interpolation functions are also supported in symbolic expressions. These functions and constants are designed to facilitate common mathematical operations and ensure compatibility with high-energy physics applications, using the \code{sympy} library.

Here are some example expressions that leverage the above functions and constants:
\begin{itemize}
    \item \code{"X > Y + log(E)"} evaluates whether $X > Y + \log(e)$.
    \item \code{"sqrt(PI) + sin(X) > 1"} checks if $\sqrt{\pi} + \sin(X) > 1$.
    \item \code{"gauss(X, 0, 1) > 0.01"} computes whether the Gaussian probability density of $X$ with mean $0$ and standard deviation $1$ is greater than $0.01$.
\end{itemize}
\begin{table*}[t]
\caption{Priors Distributions in Jarvis-HEP}
\renewcommand{\arraystretch}{1.4} 
\label{tab:priorsD}
\centering
\adjustbox{max width=\textwidth}{%
\begin{tabular}{P{1.8cm}|P{3cm}|P{9cm}}
        \Xhline{1.5pt}
        \textbf{Distribution} & \textbf{Parameters} & \textbf{Description} \\
        \Xhline{1.5pt}
        \texttt{Flat} & \code{min}, \code{max} &Uniform distribution:  $\mathcal{U}(\code{min}, \code{max})$ \\
        \texttt{Log} & \code{min}, \code{max} & Log-uniform: $\exp(\mathcal{U}(\log(\code{min}), \log(\code{max})))$ \\
        \texttt{Normal} & \code{mean}, \code{stddev} & Gaussian: $\mathcal{N}(\code{mean}, \code{stddev})$. The probability reaches 0.607 in range  $[\code{mean}-\code{stddev}, \code{mean}+\code{stddev}]$. \\
        \texttt{Log-Normal} & \code{mean}, \code{stddev} & A variable $x$ has a log-normal distribution if $\ln(x)$ is normally distributed  \\
        \texttt{Logit} & \code{location}, \code{scale} & A variable $x$ follows a logit distribution is defined as $\code{location} + \code{scale}\cdot\log(x/(1-x))$. \\
        \Xhline{1.5pt}
    \end{tabular}}
\end{table*}

 All functions are case-sensitive. Ensure that the function names and constants are written exactly as shown above. If a variable is missing in the provided dictionary during evaluation, an error will be raised.

\section{One dimensional prior distributions}
\label{app:dist}
Jarvis-HEP supports various probability distributions for sampling variables in YAML configuration files. Each variable type is specified using the \ykey{distribution} key, with associated parameters defining its behaviour.
The following example demonstrates how different variable types are defined in a Jarvis-HEP YAML file:
\begin{lstyaml}[style=yaml, caption={Example of variable definitions in YAML}]
Sampling:
  Variables:
  - name: x1
    description: "A flat distributed variable x1"
    distribution:
      type: Flat
      parameters: {min: 0, max: 10}
  
  - name: x2
    description: "A Log distributed variable x2"
    distribution:
      type: Log
      parameters: {min: 0.1, max: 10}
  	  	
  - name: x3
    description: "A Normal type variable x3"
    distribution:
      type: Normal
      parameters: {mean: 10, stddev: 5}
  	  	
  - name: x4
    description: "A Log-Normal type variable x4"
    distribution:
      type: Log-Normal
      parameters: {mean: 10, stddev: 5}
  	  	
  - name: x5
    description: "A Log-Normal type variable x5"
    distribution:
      type: Logit
      parameters: {location: 10, scale: 5}
\end{lstyaml}

\begin{itemize}
    \item Distributions are specified using the \ykey{distribution} key.
    \item \ykey{parameters} must be provided for each distribution type.
    \item Log-based distributions require positive values for bounds.
\end{itemize}

\section{Command line parsing and path resolution}
\label{app:addsyb}
\begin{table}[th]
\caption{Placeholders in path resolution}
\renewcommand{\arraystretch}{1.4} 
\label{tab:pathplhd}
\centering
\adjustbox{max width=0.49\textwidth}{%
\begin{tabular}{P{1.8cm}|P{6cm}}
\Xhline{1.5pt}
\textbf{Placeholder} 	& \textbf{Description} \\
\Xhline{1.5pt}
\code{\&J}        		& Root path of the current \code{Jarvis} project. \\
\code{{\raise.17ex\hbox{$\mathtt{\sim}$}}}        & User's home directory. \\
\code{@PackID}    		& Identifier assigned when a module enables \ykey{clone\_shadow} in the calculator. It distinguishes between multiple instances of the same module. \\
\code{@SampleID}		& The \code{uuid} code serves as a unique identifier automatically assigned to the current sampling point during operation. \\
\Xhline{1.5pt}
\end{tabular}
}
\end{table}
When writing your own YAML input file, you need to write many paths and commands. However, in the specific execution process of \code{Jarvis-HEP}, the execution of commands is slightly different from the user opening a terminal to execute commands. This difference is because \code{Jarvis-HEP} uses an \code{asyncio} subprocess to execute a shell command. 
\par Table~\ref{tab:pathplhd} defines three placeholders used in \code{Jarvis-HEP}, which can make the path in a YAML input file more compact. When executing a command list, the initial path is specified like this: 
\begin{itemize}
	\item The initial path for installing a library module is determined by the \ykey{path} option in \ykey{LibDeps}.
	\item The initial installation path for a calculator module is the factory's base directory, specified by the \ykey{path} option in the \ykey{Calculators} section. 
	\item The path for the first initialization command and/or the first execution command is given by the \ykey{path} option specified inside the module. 
\end{itemize}

\clearpage
\subsection*{~}
\addcontentsline{toc}{section}{References}
\bibliographystyle{elsarticle-num}
\bibliography{citation}

\end{document}

%% file: def.tex
\usepackage{graphicx} 
\usepackage{xcolor}
\usepackage{amsmath}
\usepackage{bm}
\usepackage{amssymb}
\usepackage{multirow}
\usepackage{makecell}
\usepackage{array}
\usepackage{siunitx} 
\usepackage{booktabs}
\usepackage[T1]{fontenc}
\usepackage{natbib}
\usepackage{algorithm}
\usepackage{algorithmic}
\usepackage{inconsolata}
\usepackage{adjustbox}
\usepackage{geometry}  
\newcolumntype{P}[1]{>{\raggedright\arraybackslash}p{#1}}
\newcolumntype{M}[1]{>{\centering\arraybackslash}m{#1}}
\newcolumntype{L}[1]{>{\raggedright\arraybackslash}m{#1}}
\newcolumntype{C}[1]{>{\centering\arraybackslash}p{#1}}
\definecolor{codegray}{HTML}{00273D}
\usepackage{lineno}
\usepackage{listings}
\usepackage{dirtree}
\usepackage{subcaption} 
\definecolor{solarized@base03}{HTML}{002B36}
\definecolor{solarized@base02}{HTML}{073642}
\definecolor{solarized@base01}{HTML}{586e75}
\definecolor{solarized@base00}{HTML}{657b83}
\definecolor{solarized@base0}{HTML}{839496}
\definecolor{solarized@base1}{HTML}{93a1a1}
\definecolor{solarized@base2}{HTML}{EEE0B5}
\definecolor{solarized@base3}{HTML}{FDF6E3}
\definecolor{solarized@yellow}{HTML}{B58900}
\definecolor{solarized@magenta}{HTML}{D33682}
\definecolor{solarized@cyan}{HTML}{2AA198}
\definecolor{solarized@orange}{HTML}{CB4B16}
\definecolor{solarized@red}{HTML}{DC322F}
\definecolor{solarized@violet}{HTML}{7F57C4}
\definecolor{solarized@blue}{HTML}{246CD2}
\definecolor{solarized@green}{HTML}{3E8700}
\definecolor{black}{rgb}{0,0,0}
\definecolor{darkred}{HTML}{550003}
\definecolor{darkgreen}{HTML}{00AA00}
\definecolor{orchid}{HTML}{AF06F5}
\definecolor{filegreen}{HTML}{128346} 
\definecolor{termframe}{HTML}{B8B8B8}
\definecolor{termbg}{HTML}{F7F7F7}
\definecolor{yamlframe}{HTML}{6FAF8A}
\definecolor{yamlbg}{HTML}{F5FBF7}
\usepackage{inconsolata}
\usepackage{hyperref}
\hypersetup{
    colorlinks=true,       
    linkcolor=violet,        
    citecolor=green,       
    filecolor=magenta,     
    urlcolor=red           
}
\newcommand{\code}[1]{\textcolor{codegray}{\texttt{#1}}}

\newcommand{\vars}{\textcolor{magenta}}

\newcommand{\file}[1]{{\color{filegreen}{\textit{#1}}}}
\usepackage{enumitem}
\newcommand{\ykey}[1]{{\color{solarized@blue}{\texttt{#1}}}}
\newcommand{\yvalue}[1]{{\color{solarized@green}{\texttt{#1}}}}

\newcommand\termplainstyle{\footnotesize\ttfamily}
\newcommand\customtilde{{\raisebox{0.2ex}{\scalebox{0.7}{\textasciitilde}}}}
\newcommand\ProcessThreeDashes{\llap{\color{solarized@red}\mdseries-{-}-}}
\newcommand\YAMLkeystyle{\color{solarized@blue}\ttfamily}
\newcommand\YAMLcommentstyle{\color{solarized@orange}\ttfamily}
\newcommand\YAMLstringstyle{\color{solarized@green}\ttfamily}
\newcommand\YAMLvaluestyle{\color{blue}\ttfamily}

\lstdefinestyle{cpp}{
  language=C++,
  basicstyle=\footnotesize\ttfamily,
  numbers=none,
  tabsize=2,
  breaklines=true,
  showstringspaces=false,
  commentstyle=\color{solarized@violet}\itshape,
  keywordstyle=\color{solarized@orange}\bfseries
}

\lstdefinestyle{terminal}
{
  language=bash,
  basicstyle=\termplainstyle,  
  numbers=none,
  tabsize=2,
  breaklines=true,
  frame=single,
  showstringspaces=false,
  numberstyle=\tiny\color{solarized@base01},
  keywordstyle=\color{solarized@orange},
  stringstyle=\color{solarized@red}\ttfamily,
  identifierstyle=\color{black},
  commentstyle=\color{solarized@violet},
  emphstyle=\color{solarized@green},
  backgroundcolor=\color{termbg},
  rulecolor=\color{termframe},
  rulesepcolor=\color{termframe},
  morekeywords={gambit, cmake, make, mkdir, gum, python, python3, wget, tar, cp, pippi, mpirun, pip, clone},
  deletekeywords={test},
  literate =
    {/gambit}{{/}{\color{black}}gambit}6
    {gambit/}{{\color{black}}gambit{/}}6
    {/include}{{/}{\color{black}}include}8
    {cmake/}{{\color{black}}cmake/}6
    {.cmake}{{.}{\color{black}}cmake}6
    {.gum}{{.}{\color{black}}gum}6
    {.tar}{{.}{\color{black}}tar}4
    {source/}{{\color{black}}source{/}}7
    { type}{{\color{black}}{}type}5
    {~}{\customtilde}1
    {math}{{\color{solarized@orange}}math}4
}

\lstnewenvironment{lstterm}{\lstset{style=terminal}}{\lstset{style=cpp}}

\lstdefinestyle{yaml}
{
  language=bash,
  keywords={true,false,null},
  otherkeywords={},
  keywordstyle=\color{solarized@base0}\bfseries,
  basicstyle=\footnotesize\color{black}\ttfamily,
  identifierstyle=\YAMLkeystyle,
  sensitive=false,
  frame=single,
  commentstyle=\color{solarized@violet},
  backgroundcolor=\color{yamlbg},
  rulecolor=\color{yamlframe},
  rulesepcolor=\color{yamlframe},
  commentstyle=\YAMLcommentstyle,
  morecomment=[l]{\#},
  morecomment=[s]{/*}{*/},
  stringstyle=\YAMLstringstyle\ttfamily,
  moredelim=**[s][\YAMLkeystyle]{,}{:},  
  moredelim=**[l][\YAMLvaluestyle]{:},  
  morestring=[b]',
  morestring=[b]",
  showspaces=false,                
  showstringspaces=false,
  breakindent=0pt,
  breakatwhitespace,
  breaklines=true,            
  breakatwhitespace=false,    
  prebreak=\raisebox{0ex}[0ex][0ex]{\textcolor{red}{$\hookleftarrow$}},
  postbreak=\raisebox{0ex}[0ex][0ex]{\textcolor{red}{$\hookrightarrow$}}, 
  columns=fullflexible,
  literate =
    {---}{{\ProcessThreeDashes}}3
    {>}{{\textcolor{solarized@red}\textgreater}}1
    {gtr}{\textgreater}1
    {grt}{\textgreater}1
    {|}{{\textcolor{solarized@red}\textbar}}1
    {\}}{{{\color{black} \}}}}1
    {\{}{{{\color{black} \{}}}1
    {[}{{{\color{black} [}}}1
    {]}{{{\color{black} ]}}}1
    {~}{\customtilde}1,
}

\lstnewenvironment{lstyaml}{\lstset{style=yaml}}{\lstset{style=cpp}}